\pdfoutput=1
\PassOptionsToPackage{table}{xcolor}
\documentclass[]{fairmeta}

\title{Watermark Basins: Black-Box Detection of Distilled LLMs}

\newcommand{\core}{\star}         %
\newcommand{\key}{\dagger}        %
\newcommand{\support}{\mathsection} %

\renewcommand\authorformat[2][]{\mbox{\sffamily\bfseries #2$^{#1}$}}

\author[\core]{Hongyan Chang}
\author[\core]{Hady Elsahar}
\author[\key]{Tom Sander}
\author[\key]{Pierre Fernandez}
\author{Saeed Mahloujifar}
\author{Sylvestre-Alvise Rebuffi}
\author{Tom\'{a}\v{s} Sou\v{c}ek}
\author{Tuan Tran}
\author{Valeriu Lacatusu}
\author[\support]{Surya Parimi}
\author[\support]{Rashel Moritz}
\author[\support]{Vanessa Stark}
\author{Alexandre Mourachko}
\affiliation{FAIR, Meta Superintelligence Labs}
\contribution[\core]{Main contributors}
\contribution[\key]{Core team}
\contribution[\support]{Project support}

\abstract{
Proprietary language models and their reasoning traces are vulnerable to API-based distillation, where an adversary trains a student model on the teacher model's outputs.
To obtain evidence of distillation, model providers can watermark the teacher's outputs, aiming to leave a radioactive signal in the student's outputs.
Existing black-box detectors search for such radioactivity using token-level statistics of the suspect model's outputs.
However, token-level signals are not always detectable, for instance when the distillation traces are sampled at a low temperature.
In this work, we discover the \emph{watermark basin}, a persistent signal that enables more reliable watermark detection: during fine-tuning, watermarked teacher traces embed a distinct, key-specific \emph{geometric signal in parameter space} that lies orthogonal to the ordinary task-specific fine-tuning subspace.
Building on this, we introduce Watermark Basin Alignment (\method), a black-box watermark detector that queries the suspect model through its API, without access to its weights or logits.
\method maps generated text to gradient representations using a fixed clean probe model, projects out the ordinary fine-tuning task directions, and tests whether the residual gradient aligns with the watermark basin induced by the watermark key.
We evaluate \method extensively across diverse teacher models and student architectures (from 135M to 8.2B parameters) distilled on mathematical reasoning.
At a strict 0.1\% false-positive rate, \method detects 79\% of distilled models at $n$-gram${=}1$ and 45\% at $n$-gram${=}3$, substantially improving over the state-of-the-art token-level baseline (29\% and 1.3\%, respectively).
Because the watermark signal resides in a persistent basin, it survives post-distillation modifications including quantization, pruning, and reinforcement-learning post-training.

}

\usepackage[utf8]{inputenc}
\usepackage{url}
\usepackage{amsfonts}
\usepackage{amssymb}
\usepackage{amsmath}
\usepackage{nicefrac}
\usepackage{dsfont}
\usepackage{bbm}
\usepackage{enumitem}
\usepackage{algorithm}
\usepackage{algpseudocode}

\usepackage{longtable}
\usepackage{siunitx}
\usepackage{rotating}
\usepackage{tabularx}
\usepackage{tabulary}
\usepackage{wrapfig}
\usepackage{listings}
\usepackage{float}
\usepackage{titletoc}   %

\newcolumntype{Y}{>{\raggedright\arraybackslash}X}

\tcbuselibrary{skins, breakable, raster}
\usetikzlibrary{decorations.pathreplacing, calc, shadows, arrows.meta, patterns}
\usepackage{pgfplots}
\usepgfplotslibrary{groupplots,statistics}
\usepackage{pgfplotstable}
\pgfplotsset{compat=1.18}

\DeclareMathSizes{4.8}{4.8}{5}{5}
\DeclareMathSizes{5.5}{5.5}{5}{5}
\DeclareMathSizes{5.8}{5.8}{5}{5}
\DeclareMathSizes{6.2}{6.2}{5}{5}
\DeclareMathSizes{6.3}{6.3}{5}{5}
\DeclareMathSizes{6.5}{6.5}{5}{5}
\DeclareMathSizes{7.5}{7.5}{6}{5}

\newcommand{\figfont}{\fontfamily{cmss}\selectfont}

\definecolor{detBBcol}{HTML}{0072B2}    %
\definecolor{detAttrcol}{HTML}{D55E00}  %
\definecolor{detReadcol}{HTML}{009E73}  %
\definecolor{detEWScol}{HTML}{E69F00}   %
\definecolor{detIdealcol}{HTML}{999999} %
\definecolor{alongcol}{HTML}{117733}    %

\definecolor{accGsmcol}{HTML}{CC79A7}   %
\definecolor{accMathcol}{HTML}{56B4E9}  %
\definecolor{accRefcol}{HTML}{9E9E9E}   %
\definecolor{accArmAcol}{HTML}{BCBDDC}  %
\definecolor{accArmBcol}{HTML}{9E9AC8}  %
\definecolor{accArmCcol}{HTML}{6A51A3}  %

\pgfplotsset{
  detBB/.style={
    color=detBBcol, mark=*, mark size=1.1pt,
    mark options={fill=detBBcol, draw=detBBcol}, line width=1.2pt},
  detAttr/.style={
    color=detAttrcol, mark=triangle*, mark size=1.5pt,
    mark options={fill=detAttrcol, draw=detAttrcol}, line width=1.2pt},
  detRead/.style={
    color=detReadcol, mark=square*, mark size=1.1pt,
    mark options={fill=detReadcol, draw=detReadcol}, line width=1.2pt},
  detEWS/.style={
    color=detEWScol, mark=diamond*, mark size=1.5pt,
    mark options={fill=detEWScol, draw=detEWScol}, line width=1.2pt},
  detIdeal/.style={
    color=detIdealcol, dashed, no marks, line width=0.9pt},
}

\definecolor{knRankcol}{HTML}{0072B2}   %
\definecolor{knEpochcol}{HTML}{D55E00}  %
\definecolor{knBatchcol}{HTML}{009E73}  %
\pgfplotsset{
  armfilter/.style={x filter/.code={\edef\thisarm{\thisrow{armid}}%
    \ifnum\thisarm=#1\else\fi}},
}

\definecolor{heatours}{HTML}{D55E00} %
\definecolor{heatbb}{HTML}{0072B2}   %
\definecolor{heatread}{HTML}{009E73} %
\colorlet{ourshue}{heatours!65!white}
\colorlet{bbhue}{heatbb!65!white}
\colorlet{readhue}{heatread!28!white}
\definecolor{detectcol}{RGB}{250,236,224}

\newcommand{\hcw}{0.50cm}                                                    %
\newcommand{\ho}[1]{\cellcolor{ourshue!#1}\makebox[\hcw][c]{#1}}             %
\newcommand{\hob}[1]{\cellcolor{ourshue!#1}\makebox[\hcw][c]{\textbf{#1}}}   %
\newcommand{\hb}[1]{\cellcolor{bbhue!#1}\makebox[\hcw][c]{#1}}               %
\newcommand{\hbb}[1]{\cellcolor{bbhue!#1}\makebox[\hcw][c]{\textbf{#1}}}     %
\newcommand{\hr}[1]{\cellcolor{readhue!#1}\makebox[\hcw][c]{#1}}             %
\newcommand{\nh}[1]{\makebox[\hcw][c]{#1}}                                   %
\newcommand{\na}{\makebox[\hcw][c]{}}                                        %

\newcommand{\heatbox}[2]{{\setlength{\fboxsep}{2pt}\colorbox{#1}{\makebox[\hcw][c]{#2}}}}
\newcommand{\hoc}[1]{\heatbox{ourshue!#1}{#1}}                               %
\newcommand{\hbc}[1]{\heatbox{bbhue!#1}{#1}}                                 %

\newcommand{\hgutb}{\hspace{4pt}}            %
\newcommand{\hgutl}{\hspace{6pt}}            %
\newcommand{\hgutg}{\hspace{10pt}}           %
\newcolumntype{N}{c @{\hgutb} *{3}{c} @{\hgutb} *{3}{c}}                %
\newcolumntype{M}{  *{3}{c} @{\hgutb} *{3}{c}}                          %
\newcolumntype{P}{c @{\hgutb} *{3}{c} @{\hgutb} *{3}{c} @{\hgutb} *{3}{c}} %
\newcolumntype{Q}{  *{3}{c} @{\hgutb} *{3}{c} @{\hgutb} *{3}{c}}        %
\newcolumntype{C}{  *{2}{c} @{\hgutb} *{2}{c}}                          %
\makeatletter
\expandafter\let\csname NC@find@W\endcsname\@undefined
\makeatother
\newcolumntype{W}{l @{\hgutl} N @{\hgutg} N @{\hgutg} N}
\newcolumntype{U}{l @{\hgutl} M @{\hgutg} M @{\hgutg} M}
\newcolumntype{J}{l @{\hgutl} C @{\hgutg} C @{\hgutg} C}                %
\newcolumntype{A}{l @{\hgutl} P @{\hgutg} P @{\hgutg} P}
\newcolumntype{B}{l @{\hgutl} Q @{\hgutg} Q @{\hgutg} Q}
\newcolumntype{G}{l @{\hgutl} N}             %
\makeatletter
\expandafter\let\csname NC@find@Y\endcsname\@undefined
\makeatother
\newcolumntype{Y}{l @{\hgutl} M}             %
\newcolumntype{T}{l @{\hgutl} *{3}{c} @{\hgutg} *{3}{c} @{\hgutg} *{3}{c}}   %
\newcolumntype{D}{l @{\hgutl} *{3}{c}}                                       %
\newcolumntype{H}{l @{\hgutl} P}             %
\newcolumntype{F}{l @{\hgutl} P @{\hgutg} P} %

\newcolumntype{R}{l @{\hgutl} c @{\hgutb} c @{\hgutb} *{3}{c} @{\hgutb} *{3}{c}}                    %
\newcolumntype{K}{l @{\hgutl} c @{\hgutb} c @{\hgutb} *{3}{c} @{\hgutb} *{3}{c} @{\hgutb} *{3}{c}}  %

\newsavebox{\housesinkbox}
\newcolumntype{E}{>{\setbox\housesinkbox=\hbox\bgroup}c<{\egroup}@{} *{3}{c} @{\hgutb} *{3}{c} @{\hgutb} *{3}{c} @{\hgutb} *{3}{c}} %
\newcolumntype{V}{l @{\hgutl} E @{\hgutg} E} %
\newcolumntype{Z}{l @{\hgutl} E}             %

\newcommand{\househead}[2][W]{\csname househeadgrid#1\endcsname{#2}}

\newcommand{\hhacc}{Acc.}
\newcommand{\hhdacc}{$\Delta$Acc.}
\newcommand{\hhdet}{\multicolumn{3}{c}{\method} & \multicolumn{3}{c}{Black-box}}
\newcommand{\hhdetr}{\hhdet & \multicolumn{3}{c}{Open-weight}}
\newcommand{\hhtpr}{\multicolumn{3}{c}{\scriptsize TPR@FPR}}
\newcommand{\hhfpr}{\nh{$5\%$} & \nh{$1\%$} & \nh{$.1\%$}}
\newcommand{\hhdetk}{\multicolumn{2}{c}{\method} & \multicolumn{2}{c}{Black-box}}
\newcommand{\hhtprk}{\multicolumn{2}{c}{\scriptsize TPR@FPR}}
\newcommand{\hhfprk}{\nh{$5\%$} & \nh{$1\%$}}
\newcommand{\hhalong}{Along $w$}
\newcommand{\hhdetow}{\hhdet & \multicolumn{6}{c}{Open-weight}}
\newcommand{\hhowsub}{\multicolumn{3}{c}{\scriptsize orig.} & \multicolumn{3}{c}{\scriptsize ideal}}

\newcommand{\househeadWbody}[2]{%
& \multicolumn{7}{c}{$n$-gram${=}1$} & \multicolumn{7}{c}{$n$-gram${=}2$} & \multicolumn{7}{c}{$n$-gram${=}3$} \\
\cmidrule(lr){2-8}\cmidrule(lr){9-15}\cmidrule(lr){16-22}
#1 & \hhacc & \hhdet & \hhacc & \hhdet & \hhacc & \hhdet \\
\cmidrule(lr){3-5}\cmidrule(lr){6-8}\cmidrule(lr){10-12}\cmidrule(lr){13-15}\cmidrule(lr){17-19}\cmidrule(lr){20-22}
& & \hhtpr & \hhtpr & & \hhtpr & \hhtpr & & \hhtpr & \hhtpr \\
& & #2 & \hhfpr & & \hhfpr & \hhfpr & & \hhfpr & \hhfpr \\
\midrule
}
\expandafter\newcommand\csname househeadgridW\endcsname[1]{\househeadWbody{#1}{\hhfpr}}

\expandafter\newcommand\csname househeadgridU\endcsname[1]{%
& \multicolumn{6}{c}{$n$-gram${=}1$} & \multicolumn{6}{c}{$n$-gram${=}2$} & \multicolumn{6}{c}{$n$-gram${=}3$} \\
\cmidrule(lr){2-7}\cmidrule(lr){8-13}\cmidrule(lr){14-19}
#1 & \hhdet & \hhdet & \hhdet \\
\cmidrule(lr){2-4}\cmidrule(lr){5-7}\cmidrule(lr){8-10}\cmidrule(lr){11-13}\cmidrule(lr){14-16}\cmidrule(lr){17-19}
& \hhtpr & \hhtpr & \hhtpr & \hhtpr & \hhtpr & \hhtpr \\
& \hhfpr & \hhfpr & \hhfpr & \hhfpr & \hhfpr & \hhfpr \\
\midrule
}

\expandafter\newcommand\csname househeadgridJ\endcsname[1]{%
& \multicolumn{4}{c}{$n$-gram${=}1$} & \multicolumn{4}{c}{$n$-gram${=}2$} & \multicolumn{4}{c}{$n$-gram${=}3$} \\
\cmidrule(lr){2-5}\cmidrule(lr){6-9}\cmidrule(lr){10-13}
#1 & \hhdetk & \hhdetk & \hhdetk \\
\cmidrule(lr){2-3}\cmidrule(lr){4-5}\cmidrule(lr){6-7}\cmidrule(lr){8-9}\cmidrule(lr){10-11}\cmidrule(lr){12-13}
& \hhtprk & \hhtprk & \hhtprk & \hhtprk & \hhtprk & \hhtprk \\
& \hhfprk & \hhfprk & \hhfprk & \hhfprk & \hhfprk & \hhfprk \\
\midrule
}

\expandafter\newcommand\csname househeadgridA\endcsname[1]{%
& \multicolumn{10}{c}{$n$-gram${=}1$} & \multicolumn{10}{c}{$n$-gram${=}2$} & \multicolumn{10}{c}{$n$-gram${=}3$} \\
\cmidrule(lr){2-11}\cmidrule(lr){12-21}\cmidrule(lr){22-31}
#1 & \hhacc & \hhdetr & \hhacc & \hhdetr & \hhacc & \hhdetr \\
\cmidrule(lr){3-5}\cmidrule(lr){6-8}\cmidrule(lr){9-11}\cmidrule(lr){13-15}\cmidrule(lr){16-18}\cmidrule(lr){19-21}\cmidrule(lr){23-25}\cmidrule(lr){26-28}\cmidrule(lr){29-31}
& & \hhtpr & \hhtpr & \hhtpr & & \hhtpr & \hhtpr & \hhtpr & & \hhtpr & \hhtpr & \hhtpr \\
& & \hhfpr & \hhfpr & \hhfpr & & \hhfpr & \hhfpr & \hhfpr & & \hhfpr & \hhfpr & \hhfpr \\
\midrule
}

\expandafter\newcommand\csname househeadgridG\endcsname[1]{%
#1 & \hhacc & \hhdet \\
\cmidrule(lr){3-5}\cmidrule(lr){6-8}
& & \hhtpr & \hhtpr \\
& & \hhfpr & \hhfpr \\
\midrule
}

\expandafter\newcommand\csname househeadgridY\endcsname[1]{%
#1 & \hhdet \\
\cmidrule(lr){2-4}\cmidrule(lr){5-7}
& \hhtpr & \hhtpr \\
& \hhfpr & \hhfpr \\
\midrule
}

\expandafter\newcommand\csname househeadgridT\endcsname[1]{%
& \multicolumn{3}{c}{$n$-gram${=}1$} & \multicolumn{3}{c}{$n$-gram${=}2$} & \multicolumn{3}{c}{$n$-gram${=}3$} \\
\cmidrule(lr){2-4}\cmidrule(lr){5-7}\cmidrule(lr){8-10}
#1 & \hhtpr & \hhtpr & \hhtpr \\
& \hhfpr & \hhfpr & \hhfpr \\
\midrule
}

\expandafter\newcommand\csname househeadgridD\endcsname[1]{%
& \multicolumn{3}{c}{$n$-gram${=}1$} \\
\cmidrule(lr){2-4}
#1 & \hhtpr \\
& \hhfpr \\
\midrule
}

\expandafter\newcommand\csname househeadgridF\endcsname[1]{%
& \multicolumn{10}{c}{$n$-gram${=}1$} & \multicolumn{10}{c}{$n$-gram${=}2$} \\
\cmidrule(lr){2-11}\cmidrule(lr){12-21}
#1 & \hhdacc & \hhdetr & \hhdacc & \hhdetr \\
\cmidrule(lr){3-5}\cmidrule(lr){6-8}\cmidrule(lr){9-11}\cmidrule(lr){13-15}\cmidrule(lr){16-18}\cmidrule(lr){19-21}
& & \hhtpr & \hhtpr & \hhtpr & & \hhtpr & \hhtpr & \hhtpr \\
& & \hhfpr & \hhfpr & \hhfpr & & \hhfpr & \hhfpr & \hhfpr \\
\midrule
}

\expandafter\newcommand\csname househeadgridH\endcsname[1]{%
#1 & \hhdacc & \hhdetr \\
\cmidrule(lr){3-5}\cmidrule(lr){6-8}\cmidrule(lr){9-11}
& & \hhtpr & \hhtpr & \hhtpr \\
& & \hhfpr & \hhfpr & \hhfpr \\
\midrule
}

\expandafter\newcommand\csname househeadgridR\endcsname[1]{%
#1 & \hhacc & \hhalong & \hhdet \\
\cmidrule(lr){4-6}\cmidrule(lr){7-9}
& & & \hhtpr & \hhtpr \\
& & & \hhfpr & \hhfpr \\
\midrule
}
\expandafter\newcommand\csname househeadgridK\endcsname[1]{%
#1 & \hhdacc & \hhalong & \hhdetr \\
\cmidrule(lr){4-6}\cmidrule(lr){7-9}\cmidrule(lr){10-12}
& & & \hhtpr & \hhtpr & \hhtpr \\
& & & \hhfpr & \hhfpr & \hhfpr \\
\midrule
}

\expandafter\newcommand\csname househeadgridV\endcsname[1]{%
& \multicolumn{13}{c}{$n$-gram${=}2$} & \multicolumn{13}{c}{$n$-gram${=}3$} \\
\cmidrule(lr){2-14}\cmidrule(lr){15-27}
#1 & \hhacc & \hhdetow & \hhacc & \hhdetow \\
\cmidrule(lr){3-5}\cmidrule(lr){6-8}\cmidrule(lr){9-14}\cmidrule(lr){16-18}\cmidrule(lr){19-21}\cmidrule(lr){22-27}
& & \hhtpr & \hhtpr & \hhowsub & & \hhtpr & \hhtpr & \hhowsub \\
& & \hhfpr & \hhfpr & \hhfpr & \hhfpr & & \hhfpr & \hhfpr & \hhfpr & \hhfpr \\
\midrule
}

\newcommand{\househeadZbody}[2]{%
& \multicolumn{13}{c}{#2} \\
\cmidrule(lr){2-14}
\househeadZstack{#1}%
}
\newcommand{\househeadZstack}[1]{%
#1 & \hhacc & \hhdetow \\
\cmidrule(lr){3-5}\cmidrule(lr){6-8}\cmidrule(lr){9-14}
& & \hhtpr & \hhtpr & \hhowsub \\
& & \hhfpr & \hhfpr & \hhfpr & \hhfpr \\
\midrule
}
\expandafter\newcommand\csname househeadgridZS\endcsname[1]{\househeadZstack{#1}}
\expandafter\newcommand\csname househeadgridZ1\endcsname[1]{\househeadZbody{#1}{$n$-gram${=}1$}}
\expandafter\newcommand\csname househeadgridZ2\endcsname[1]{\househeadZbody{#1}{$n$-gram${=}2$}}
\expandafter\newcommand\csname househeadgridZ3\endcsname[1]{\househeadZbody{#1}{$n$-gram${=}3$}}

\expandafter\newcommand\csname househeadgridB\endcsname[1]{%
& \multicolumn{9}{c}{$n$-gram${=}1$} & \multicolumn{9}{c}{$n$-gram${=}2$} & \multicolumn{9}{c}{$n$-gram${=}3$} \\
\cmidrule(lr){2-10}\cmidrule(lr){11-19}\cmidrule(lr){20-28}
#1 & \hhdetr & \hhdetr & \hhdetr \\
\cmidrule(lr){2-4}\cmidrule(lr){5-7}\cmidrule(lr){8-10}\cmidrule(lr){11-13}\cmidrule(lr){14-16}\cmidrule(lr){17-19}\cmidrule(lr){20-22}\cmidrule(lr){23-25}\cmidrule(lr){26-28}
& \hhtpr & \hhtpr & \hhtpr & \hhtpr & \hhtpr & \hhtpr & \hhtpr & \hhtpr & \hhtpr \\
& \hhfpr & \hhfpr & \hhfpr & \hhfpr & \hhfpr & \hhfpr & \hhfpr & \hhfpr & \hhfpr \\
\midrule
}

\newsavebox{\housebx}
\newlength{\housew}
\newcommand{\housetable}[1]{%
  \sbox{\housebx}{#1}%
  \setlength{\housew}{\wd\housebx}%
  \ifdim\housew>\linewidth \setlength{\housew}{\linewidth}\fi
  \resizebox{\housew}{!}{\usebox{\housebx}}%
}

\newcommand{\housematchscale}{0.8465}
\newcommand{\housetablematch}[1]{%
  \sbox{\housebx}{#1}%
  \setlength{\housew}{\housematchscale\wd\housebx}%
  \resizebox{\housew}{!}{\usebox{\housebx}}%
}

\newcommand{\housesetup}{%
  \footnotesize
  \setlength{\tabcolsep}{2pt}%
  \renewcommand{\arraystretch}{1.1}%
}

\usepackage{amsthm}
\usepackage{ifthen}
\newboolean{set_me_to_remove_todos}
\setboolean{set_me_to_remove_todos}{true} %
\ifdefined\hidecomments\setboolean{set_me_to_remove_todos}{true}\fi
\newcommand{\todo}[1]{{\color{red} [\textbf{TODO}: #1]}}
\newcommand{\pierre}[1]{{\color{blue} [\textbf{Pierre}: #1]}}
\newcommand{\hongyan}[1]{{\color{orange} [\textbf{Hongyan}: #1]}}
\newcommand{\tom}[1]{{\color{red} [\textbf{Tom}: #1]}}
\newcommand{\hady}[1]{{\color{purple} [\textbf{Hady}: #1]}}
\newcommand{\alex}[1]{{\color{teal} [\textbf{Alex}: #1]}}
\newcommand{\sylvestre}[1]{{\color{olive} [\textbf{Sylvestre}: #1]}}
\newcommand{\tomas}[1]{{\color{brown} [\textbf{Tomas}: #1]}}
\newcommand{\tuan}[1]{{\color{magenta} [\textbf{Tuan}: #1]}}
\newcommand{\valeriu}[1]{{\color{cyan} [\textbf{Valeriu}: #1]}}
\newcommand{\claude}[1]{{\color{violet} [\textbf{Claude}: #1]}}

\definecolor{ceditpink}{HTML}{D81B60}

\definecolor{aeditgreen}{HTML}{1B7F3B}

\definecolor{leditamber}{HTML}{8C5A00}

\ifthenelse{\boolean{set_me_to_remove_todos}}{
    \renewcommand{\todo}[1]{}
    \renewcommand{\pierre}[1]{}
    \renewcommand{\hongyan}[1]{}
    \renewcommand{\hady}[1]{}
    \renewcommand{\tom}[1]{}
    \renewcommand{\alex}[1]{}
    \renewcommand{\sylvestre}[1]{}
    \renewcommand{\tomas}[1]{}
    \renewcommand{\tuan}[1]{}
    \renewcommand{\valeriu}[1]{}
    \renewcommand{\claude}[1]{}
}{}

\definecolor{badred}{rgb}{0.8,0.2,0.2}
\definecolor{basecolor}{HTML}{E63946}
\definecolor{ftcolor}{HTML}{457B9D}
\definecolor{ftaugcolor}{HTML}{2A9D8F}
\definecolor{ftaugsynccolor}{HTML}{e2b33c}
\definecolor{textonlyorange}{HTML}{F4A261}
\definecolor{algcmtcol}{HTML}{4B6A88}  %

\definecolor{pkcalcol}{HTML}{56B4E9} %
\definecolor{pkhotcol}{HTML}{D55E00} %
\definecolor{pkposcol}{HTML}{009E73} %

\usepackage{pifont}
\usepackage{xspace}
\newcommand{\cmark}{\ding{51}\xspace}%
\newcommand{\xmark}{\ding{55}\xspace}%

\newcommand{\sk}{k}

\def\1{\mathbf{1}}

\def\Hnull{H_0}

\usepackage{xspace} %

\newcommand{\method}{\mbox{WM-Basin}\xspace}          %
\newcommand{\methodfull}{Watermark Basin Alignment\xspace} %
\newcommand{\probe}{\theta_{\mathrm{probe}}}         %

\newcommand{\dkey}{\Delta_{\mathrm{total}}}          %
\newcommand{\dkeyperp}{\Delta_{\mathrm{wm}}}          %
\newcommand{\dtask}{\Delta_{\mathrm{task}}}           %
\newcommand{\dwmhat}{\hat{\Delta}_{\mathrm{wm}}}      %
\newcommand{\dtaskhat}{\hat{\Delta}_{\mathrm{task}}} %
\newcommand{\dsusp}{\Delta_{\mathrm{sus}}}          %
\newcommand{\dwm}[1]{\Delta_{#1}}                    %
\newcommand{\taskbasis}{U}                           %
\newcommand{\tdsusp}{\widetilde{\Delta}_{\mathrm{sus}}} %
\newcommand{\tdwm}[1]{\widetilde{\Delta}_{#1}}       %
\newcommand{\basinscore}{S}                          %
\newcommand{\prank}{p_{\mathrm{rank}}}               %

\newcommand{\Xtrain}{\mathcal{X}_{\mathrm{train}}}   %
\newcommand{\Xbuild}{\mathcal{X}_{\mathrm{build}}}   %
\newcommand{\Xtest}{\mathcal{X}_{\mathrm{test}}}     %

\newcommand{\myparagraph}[1]{\par\vspace{0em}\noindent\textbf{#1}\hspace{0.2em}}

\newcommand{\eg}{e.g.,\@ }
\newcommand{\ie}{i.e.,\@ }

\newcolumntype{x}[2]{S[table-format=#1.#2,table-auto-round]}

\usepackage{tikz}

\usetikzlibrary{shapes.misc, positioning, decorations.pathreplacing}
\usetikzlibrary{positioning, backgrounds, fit, calc}

\definecolor{green1}{HTML}{008000}
\definecolor{green2}{HTML}{03ad55}
\definecolor{red1}{HTML}{FF0000}
\definecolor{red2}{HTML}{880000}

\lstdefinestyle{mystyle}{
    breaklines=true,
    basicstyle=\scriptsize\sffamily,
    numbers=none,
    language={},
    framextopmargin=0pt,
    framexbottommargin=0pt,
    breakindent=0pt,
    showspaces = false,
    keywordstyle=\bfseries,
    showstringspaces=false,
    columns=fullflexible,
    morekeywords={acoustic, guitar, thatch},
    moredelim=**[is][\color{green1}]{@}{@},
    moredelim=**[is][\color{red2}]{^}{^}
}

\newtcblisting{prompt}[1][]{
    arc=3pt, outer arc=3pt,
    width=.93\linewidth,
    left=1mm,
    top=0mm,
    bottom=0mm,
    title=\userheader,
    colback=red!5!white,
    colframe=red!50!black,
    fonttitle=\bfseries,
    listing only,
    listing options={style=mystyle,deletekeywords={}},
    breakable,
    #1
}

\definecolor{oai}{HTML}{10a37f}

\crefname{assumption}{Assumption}{Assumptions}
\Crefname{assumption}{Assumption}{Assumptions}

\newif\ifexternalize
\externalizefalse
\ifexternalize
  \usetikzlibrary{external}
\else
  \providecommand{\tikzexternaldisable}{}
  \providecommand{\tikzexternalenable}{}
\fi

\begin{document}
\sisetup{text-series-to-math = true}

\tikzexternaldisable
\maketitle
\tikzexternalenable

\begin{figure}[b!]
  \centering
  \includegraphics[width=\linewidth]{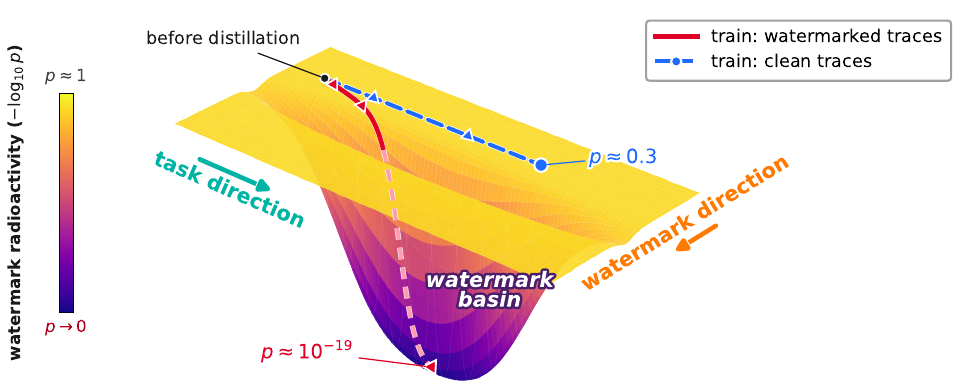}
  \caption{\textbf{The watermark basin.}
  Training on watermarked traces moves the student into a persistent, key-specific \emph{watermark basin}, whereas clean traces from the same teacher move it only along the task direction. \methodfull{} (\method) recovers this geometry from black-box outputs alone to detect distillation.}
  \label{fig:teaser}
\end{figure}

\section{Introduction}
\label{sec:introduction}

Proprietary language models are increasingly targeted by API-based distillation: an adversary queries a closed teacher model and fine-tunes a competing student model on its outputs.
This attack extends black-box model extraction, first demonstrated against ML-as-a-service systems~\citep{tramer2016stealing}, and has become a major threat to reasoning LLMs~\citep{anthropic2026distillation,gtig2026distillation,remler2026adversarial}.
Providers have responded by withholding reasoning traces, but concealment does not prevent distillation.
Hidden traces can be extracted at scale from encrypted response blocks~\citep{panfilov2026stealing}, while answers and short summaries can be expanded into synthetic traces that recover much of the benefit of training on the originals~\citep{zhang2026steal}.
Other defenses make extraction harder rather than detectable~\citep{anthropic2026distillation,gtig2026distillation,savani2026antidistillation,ma2026protecting}.
We review related work in full in \S\ref{sec:related}.

In this work, we focus on output watermarking, which is encouraged by emerging regulatory requirements~\citep{EUCodeOfPractice2026}.
Anthropic, for example, embeds watermarks in texts generated by Claude models released beginning in August 2026~\citep{anthropic2026marking}.
Such watermarks can survive distillation: \citet{sander2024watermarking} show that students trained on watermarked teacher outputs retain a ``radioactive'' signal.
Their black-box detector, to our knowledge the only existing one for this setting and our baseline throughout, tests suspect outputs for an excess of watermark-favored tokens, but only where watermark-conditioned contexts recur.
We find that this dependence is fragile: detection degrades sharply with larger context windows and loses most of its power under low-temperature teacher sampling.
\emph{We therefore seek a watermark signal that does not require exact token-level recurrence.}

\begin{figure*}[t]
  \centering
  \input{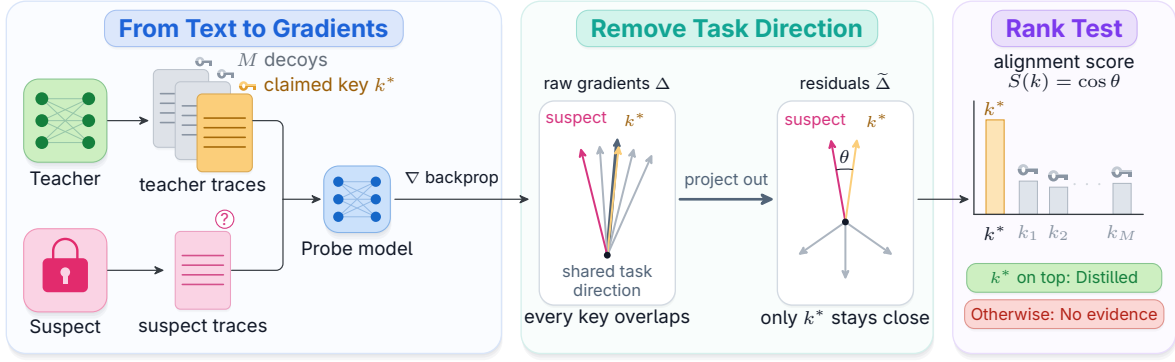}
  \caption{\textbf{\method: black-box distillation detection via watermark basin alignment.}
  The three panels are the three stages of \S\ref{subsec:closed-box}.
  \textbf{(i)}~The verifier collects suspect outputs on test prompts and, on separate build prompts, teacher outputs under the claimed watermark key and $M$ decoy keys. Backpropagating each collection through one fixed probe model gives the suspect gradient $\dsusp$ and one reference gradient $\dwm{k}$ per key.
  \textbf{(ii)}~Raw gradients share large common task components that inflate the cosine similarities, so every key appears to overlap with the suspect. Projecting out the shared task directions leaves the residuals $\tdsusp$ and $\tdwm{k}$, among which only the claimed key stays close to the suspect; $\theta$ is the angle between them.
  \textbf{(iii)}~Each candidate key receives the alignment score $\basinscore(k)=\cos(\tdsusp,\tdwm{k})$, and the claimed key's rank among the decoys gives the $p$-value $\prank$. The claimed key ranking at the top is evidence of distillation; a decoy key at the top is not.}
  \label{fig:blackbox-pipeline}
\end{figure*}

Our main finding is the \textbf{watermark basin}: supervised fine-tuning on clean (watermark-free) teacher traces moves a student along a broad \emph{task direction}, whereas watermarked traces add a faint, key-specific \emph{watermark direction} (Fig.~\ref{fig:teaser}), leading the student into a key-specific region of parameter space.
Post-training updates cannot easily push the student out of this basin.
Building on this geometry, we introduce \textbf{\methodfull (\method)}, a black-box detector that represents the suspect model's outputs and the watermarked teacher's outputs as gradients of one fixed, clean probe model.
After removing the task subspace estimated from clean models, the distilled student should align with the watermark key \emph{more closely} than with decoy keys that were never used to watermark the teacher.
\method turns this relative rank into a calibrated $p$-value using a non-parametric ranking test~\citep{dwass1957modified}: a small value is evidence that the suspect was distilled from the claimed teacher and key on the audited domain (Fig.~\ref{fig:blackbox-pipeline}).

We evaluate \method along many axes: the student architecture, the teacher model, the watermarking scheme, the sampling temperature and watermarked fraction of the distillation traces, the fine-tuning recipe, and the task domain.
Among $1{,}200$ student models distilled from DeepSeek-R1-Distill-Qwen-14B~\citep{guo2025deepseek} on OpenR1-Math reasoning traces~\citep{openr1}, at a $0.1\%$ false-positive rate, \method detects $79\%$ of distilled models at $n$-gram${=}1$ and $45\%$ at $n$-gram${=}3$, against $29\%$ and $1.3\%$ for the state-of-the-art black-box detector~\citep{sander2024watermarking}.
Quantization, pruning, and reinforcement-learning post-training all leave detection intact, as none of them moves the student out of the watermark basin.
Like any black-box method, \method reads only the suspect model's text, so detection depends on that text carrying the teacher's token distribution.
Continued training on clean data can reshape that distribution and therefore degrade black-box detection, although the basin itself survives in the weights and remains easy to identify under white-box access.
Our contributions are as follows:
\begin{itemize}[leftmargin=*,nosep,itemsep=2pt]
\item \textbf{Discovery of watermark basins.} We find that watermarked distillation places every distilled student in a persistent basin, spanned by the ordinary task direction and a faint, key-specific watermark direction. Leaving this basin without knowledge of the key is difficult: post-training operations (quantization, pruning, supervised fine-tuning, RL) displace weights primarily along the task axis and leave the key-specific component intact, which we exploit for detection.
\item \textbf{A parameter-space black-box detector (\method).} \method looks for distillation evidence in parameter space. It maps text outputs to gradients of a fixed clean probe model, removes the shared task subspace, and scores the residual alignment against decoy keys. The watermark becomes far more detectable than in token space, with exact finite-sample false-positive control and no access to the suspect's parameters.
\item \textbf{Robust detection at scale.} Across $24$ student architectures ($135$M to $8.2$B parameters), $4$ teachers, and $4$ watermarking schemes, \method achieves $79\%$ TPR at $0.1\%$ FPR versus $29\%$ for the state-of-the-art token-level detector, and remains effective under quantization, pruning, and RL post-training.
\end{itemize}

\section{Watermark-Basin Detection}
\label{sec:detection}

\subsection{Threat Model and Statistical Setup}
\label{sec:setup}

\myparagraph{Distillation setting and adversary model.}
A proprietary teacher model $\mathcal{M}_{\text{teacher}}$ serves completions watermarked under a secret key $\sk^\star\leftarrow\mathsf{KeyGen}$, drawn from the watermark scheme's key-generation procedure.
Following deployed systems~\citep{aaronson2022gumbelmax,dathathri2024scalable,sander2026textseal,anthropic2026marking}, we consider distortion-free schemes: the teacher's marginal next-token distribution is preserved, but for each prompt $x$ the output $y \sim \mathcal{M}_{\text{teacher}}(x; \sk^\star)$ is coupled to the pseudorandomness induced by $\sk^\star$.
The adversary adaptively chooses $N$ prompts $\Xtrain=\{x_i\}_{i=1}^{N}$, receives watermarked traces $\{y_i\}$, builds $\mathcal{D}=\{(x_i,y_i)\}$, and trains a student $\mathcal{M}_{\text{sus}}$ by supervised fine-tuning on $\mathcal{D}$.
Without knowledge of $\sk^\star$, watermarked completions are indistinguishable from clean text, so the default adversary is unaware of the watermark and uses the traces as-is.
After fine-tuning, the adversary may further post-train $\mathcal{M}_{\text{sus}}$ on clean data, \eg via continued SFT or GRPO~\citep{guo2025deepseek}.

\myparagraph{Verifier regime.}
An audit is opened by the verifier, either the model owner or an authorized third party, and begins by naming one capability the owner sells and the suspect appears to compete on.
We call it the \emph{audited domain} $D$ (\eg math) and let $P_D$ be its induced prompt distribution; the verifier chooses both and draws every prompt it uses from $P_D$.
It holds two disjoint pools: a \emph{build} pool $\Xbuild$ used to construct the detector, and a \emph{test} pool $\Xtest$ used to read the suspect.
Neither overlaps the adversary's prompts, which the verifier never sees, and keeping build and test prompts apart avoids score inflation from prompt reuse.
The verifier's access to $\mathcal{M}_{\text{sus}}$ is \emph{black-box}: it observes only texts sampled on the test prompts.
The access is also \emph{unsupervised} in the sense of \citet{sander2024watermarking}: the verifier does not hold the traces $\mathcal{M}_{\text{sus}}$ was trained on, only fresh teacher traces on the build prompts (unwatermarked, watermarked under $\sk^\star$, and watermarked under $M$ independent decoy keys $\sk_j\leftarrow\mathsf{KeyGen}$), together with a pool of public clean checkpoints.
It knows the system prompts used to query $\mathcal{M}_{\text{sus}}$ and the teacher, because it chooses them, and the number of test prompts sent to the suspect is the query budget.

\myparagraph{Statistical objective.}
Let $\basinscore(\sk)$ be any scalar alignment score computed from the suspect's completions on test prompts and from teacher traces on build prompts under key $\sk$; \S\ref{subsec:closed-box} instantiates it.
Its absolute magnitude is not comparable across suspects, prompt pools, or probe models, so an audit reads only where the claimed key $\sk^\star$ ranks among decoy keys.
Our null hypothesis $\Hnull$ is that the suspect's outputs on the audited domain $D$ contain no information about the production key $\sk^\star$.
If the texts do depend on $\sk^\star$, the dependence can only arise through the distillation path $\sk^\star \to \{y_i\} \to \mathcal{M}_{\text{sus}}$.
Rejecting $\Hnull$ therefore establishes \emph{watermark radioactivity}: the suspect was trained on the teacher's keyed text, which separates it from any model not distilled under $\sk^\star$.
A failure to reject is scoped to domain $D$; we measure detection power beyond the distillation domain in \S\ref{subsec:wild-access}.

\subsection{Watermark Basins in Parameter Space}
\label{subsec:direction-signal}

We first use a white-box setting to \emph{investigate} what distinguishes a student trained on watermarked traces from one trained on clean traces, and thus what makes a model radioactive.
Fig.~\ref{fig:teaser} presented the watermark basin; here we test that geometry directly with access to the student weights.
We fix the true key $\sk^\star$ and fine-tune the same base model either on teacher traces watermarked with $\sk^\star$, giving $\boldsymbol{\theta}_{\mathrm{student}}$, or on unwatermarked traces, giving a clean control $\boldsymbol{\theta}_{\mathrm{ctrl}}$.
The teacher, prompts, initialization, and training recipe are otherwise identical, so both students learn the same task while only one sees the watermark.
Let $\dkey:=\boldsymbol{\theta}_{\mathrm{student}}-\boldsymbol{\theta}_{\mathrm{base}}$ be the total update from watermarked training and $\dtask:=\boldsymbol{\theta}_{\mathrm{ctrl}}-\boldsymbol{\theta}_{\mathrm{base}}$ the clean task update.
We isolate the watermark-specific part of the update by removing from $\dkey$ its component along $\dtask$:
\begin{equation}
    \dkeyperp
    := \dkey - \frac{\langle \dkey,\dtask\rangle}{\|\dtask\|_2^2}\,\dtask .
    \label{eq:watermark_direction}
\end{equation}
The subtracted term is the part of the watermarked update that the clean student also makes, \ie learning the task.
What remains, $\dkeyperp$, is the part that only the watermark can explain.
We normalize $\dkeyperp$ and $\dtask$ to obtain the watermark axis $\dwmhat$ and the task axis $\dtaskhat$.

\begin{figure}[t]
  \centering
  \includegraphics[width=\linewidth]{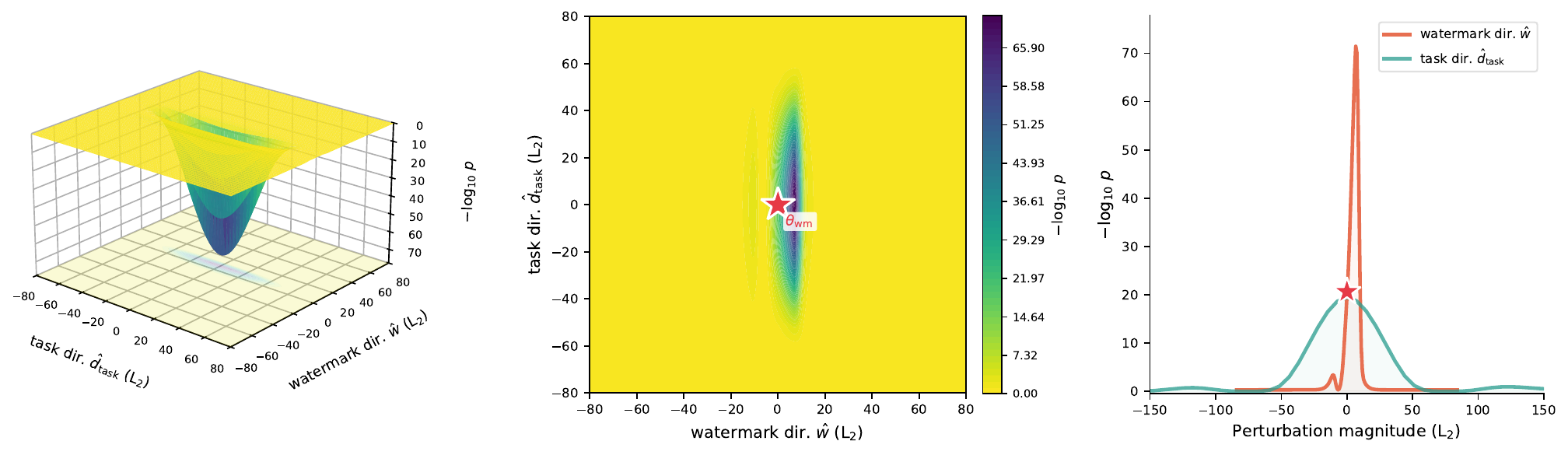}
  \caption{\textbf{The anisotropic geometry of the watermark basin}, measured as the white-box detection significance ($-\log_{10} p$) around a radioactive student $\boldsymbol{\theta}_{\mathrm{wm}}$ (red star; $\boldsymbol{\theta}_{\mathrm{student}}$ in the text) fine-tuned on watermarked teacher traces.
  \textbf{Left to right:} 3D landscape, 2D contour map, and 1D slices through $\boldsymbol{\theta}_{\mathrm{wm}}$.
  The signal falls to the detector's floor within a few steps along the watermark direction $\dwmhat$ (red curve), while remaining detectable over a much broader range along the task direction $\dtaskhat$ (green curve).}
  \label{fig:anisotropic-basin}
\end{figure}

To map the basin, we start at $\boldsymbol{\theta}_{\mathrm{student}}$ and perturb the weights along each axis, $\boldsymbol{\theta}_{\mathrm{student}} + t_{\mathrm{wm}}\dwmhat + t_{\mathrm{task}}\dtaskhat$, over a grid of step sizes $(t_{\mathrm{wm}}, t_{\mathrm{task}})$.
At each point, we apply the open-weight detector of \citet{sander2024watermarking} to the perturbed student and record $-\log_{10}p$, where a smaller $p$-value is stronger evidence of distillation.
Fig.~\ref{fig:anisotropic-basin} shows that the surface is strongly anisotropic: a small step along the watermark axis makes the watermark signal collapse, whereas much larger displacements along the task axis leave it detectable.
The basin is broad along $\dtaskhat$ but has a sharp boundary along $\dwmhat$.

This asymmetry gives the two axes different roles.
The width along $\dtaskhat$ suggests why ordinary post-training can alter task behavior without erasing the basin, a hypothesis we test under compression and continued training in \S\ref{subsec:persistence}.
The narrowness along $\dwmhat$ makes the basin key-specific, which motivates comparing against decoy keys rather than thresholding a generic watermark score.
This analysis relies on access to the student's parameters, which a black-box verifier does not have.
Next, we show how to detect the same key-specific signal using only text sampled from the suspect.

\subsection{\method: Black-Box Estimation via Probe Gradients}
\label{subsec:closed-box}
A black-box verifier cannot inspect the suspect's parameters and observes only its generated text.
\method maps suspect, keyed-teacher, and clean-model outputs into a shared gradient representation, removes the variation common to clean models, and ranks the claimed key by the remaining alignment.
Fig.~\ref{fig:blackbox-pipeline} gives the overview and Alg.~\ref{alg:wm-basin} the full procedure.

\myparagraph{Stage (i): from text to probe gradients.}
We use a small clean model as the probe, warm its LoRA parameters $\probe$ on held-out unwatermarked teacher traces, and then freeze them.
A prompt--completion pair $(x,y)$ is represented by the gradient $g(x,y)=\nabla_{\probe}\log p_{\probe}(y\mid x)$, the parameter nudge that would make the probe more likely to produce $y$ given $x$.
We average these gradients over the suspect's completions $\hat y_x$ on the test prompts, and over the teacher's completions $y^{\sk}_x$ under each key $\sk$ on the build prompts:
\begin{equation}
    \dsusp = \frac{1}{|\Xtest|}\sum_{x\in\Xtest} g\big(x,\hat y_x\big),
    \qquad
    \dwm{\sk} = \frac{1}{|\Xbuild|}\sum_{x\in\Xbuild} g\big(x,y^{\sk}_x\big).
    \label{eq:probe-gradients}
\end{equation}
If the suspect inherited key $\sk$, the nudges induced by its outputs should share the key-specific component of the nudges induced by teacher outputs under $\sk$; this alignment is the signal we measure.
The gradients give a common coordinate system for comparing first-order fine-tuning tendencies; they do not reconstruct the suspect's hidden weight update~\citep{xia2024less}.
A single fixed random projection maps every gradient to $d{=}8{,}192$ dimensions while approximately preserving inner products~\citep{johnson1984extensions}.
This stage needs only keyed teacher traces, not the watermark scheme's own detector.

\myparagraph{Stage (ii): removing the shared task subspace.}
\phantomsection\label{subsec:task-removal}
Raw probe gradients share large components caused by the task, the response format, and generic language modeling.
These common components inflate every cosine similarity $\cos(\dsusp,\dwm{\sk})$ and hide the small differences that distinguish one key from another.
We therefore adapt the \emph{All-but-the-Top} post-processing used to dehub learned representations~\citep{mu2017all}.
Clean checkpoints generate responses on held-out build prompts, and PCA on their probe gradients gives the top $q$ directions of clean-model variation, collected in the orthonormal basis $\taskbasis=[u_1,\dots,u_q]\in\mathbb{R}^{d\times q}$.
We project every gradient off this basis and score each key by the cosine of the residuals:
\begin{equation}\label{eq:residual-score}
    \widetilde\Delta=(I-\taskbasis\taskbasis^\top)\,\Delta,
    \qquad
    \basinscore(\sk)=\cos\!\big(\tdsusp,\tdwm{\sk}\big).
\end{equation}
The projection keeps only what clean models do \emph{not} have in common, so the cosine asks whether the suspect's residual points the same way as key $\sk$'s residual.
We fit $\taskbasis$ on clean models alone, never on suspects, and use $q{=}100$ (App.~\ref{app:calib-task-removal}).

\myparagraph{Stage (iii): ranking against decoy keys.}
\phantomsection\label{subsec:key-cal}
Because $\basinscore$ has no absolute scale, the verifier evaluates it at the claimed key and at $M$ independent decoy keys $\sk_1,\dots,\sk_M\leftarrow\mathsf{KeyGen}$, and reads the claimed key's rank:
\begin{equation}\label{eq:rank-pvalue}
\prank=\frac{1+\sum_{j=1}^M\mathbb{I}\!\left(\basinscore(\sk_j)\geq\basinscore(\sk^\star)\right)}{M+1}.
\end{equation}
In words, $\prank$ is the fraction of the $M{+}1$ candidate keys, the claimed key included, that score at least as high as the claimed key.
Under $\Hnull$, the claimed key and the decoys are independent outputs of the same key generator and pass through the same probe, random projection, basis $\taskbasis$, and prompts.
Their scores are therefore exchangeable, which gives $\prank$ exact finite-sample validity marginally over the draw of $\sk^\star$~\citep{dwass1957modified,pesarin2010permutation} (App.~\ref{app:key-cal}).
Under $\Hnull$ and absent ties, $\prank$ is uniform on the grid $\{1,\dots,M{+}1\}/(M{+}1)$; a very small $\prank$ lets the verifier conclude distillation from $\sk^\star$ with high confidence.
The smallest attainable value is $1/(M+1)$, so certifying $\alpha=0.01$ or $0.001$ requires $M\geq99$ or $999$ decoys.

\begin{algorithm}[t]
\small
\caption{\method: black-box detection of watermark distillation via basin alignment.}
\label{alg:wm-basin}
\begin{algorithmic}[1]
\Require
    Query access to the suspect model $\mathcal{M}_{\text{sus}}$ (black-box API) and the watermarked teacher;
    claimed key $\sk^\star$ and $M$ decoy keys $\sk_1, \dots, \sk_M \leftarrow \mathsf{KeyGen}$;
    disjoint build and test prompt pools $\Xbuild$, $\Xtest$ from the audited domain $D$;
    a pool of clean off-the-shelf models;
    significance level $\alpha$.
\Ensure $p$-value $\prank$; report the suspect as distilled into the basin of $\sk^\star$ when $\prank \le \alpha$.
\Statex
\Statex \textbf{Build once} \textcolor{blue}{\emph{(offline; independent of the suspect)}}
\State $\probe \gets$ clean small model warmed by a short LoRA pass on unwatermarked teacher traces from $\Xbuild$
\For{each key $\sk \in \{\sk^\star, \sk_1, \dots, \sk_M\}$}
    \State Sample teacher responses under key $\sk$ on $\Xbuild$
    \State $\dwm{\sk} \gets$ average probe gradient of these responses \Comment{Eq.~\eqref{eq:probe-gradients}}
\EndFor
\State $\taskbasis \gets$ top-$q$ principal directions of the clean pool's probe gradients on $\Xbuild$
\For{each key $\sk \in \{\sk^\star, \sk_1, \dots, \sk_M\}$}
    \State $\tdwm{\sk} \gets (I - \taskbasis\taskbasis^{\top})\dwm{\sk}$ \Comment{Residual reference gradient}
\EndFor
\Statex
\Statex \textbf{Detect} \textcolor{blue}{\emph{(online; once per suspect)}}
\State Sample answers from $\mathcal{M}_{\text{sus}}$ on $\Xtest$ \Comment{Unused by every step above}
\State $\dsusp \gets$ average probe gradient of the suspect's answers \Comment{Eq.~\eqref{eq:probe-gradients}}
\State $\tdsusp \gets (I - \taskbasis\taskbasis^{\top})\dsusp$ \Comment{Residual suspect gradient}
\For{each key $\sk \in \{\sk^\star, \sk_1, \dots, \sk_M\}$}
    \State $\basinscore(\sk) \gets \cos\left(\tdsusp,\, \tdwm{\sk}\right)$ \Comment{Eq.~\eqref{eq:residual-score}}
\EndFor
\State $\prank \gets \frac{1 + \sum_{j=1}^M \mathbb{I}\left(\basinscore(\sk_j) \ge \basinscore(\sk^\star)\right)}{M + 1}$ \Comment{Eq.~\eqref{eq:rank-pvalue}}
\State \Return $\prank$; report ``distilled into the basin of $\sk^\star$'' if $\prank \le \alpha$
\end{algorithmic}
\end{algorithm}

\myparagraph{Build once, audit many suspects.}
Alg.~\ref{alg:wm-basin} separates what a verifier computes once from what it pays per suspect.
The \emph{build-once} block never touches the suspect: it reads only the candidate keys, the build pool $\Xbuild$, and the clean checkpoints, and produces the warmed probe, the reference bank $\{\tdwm{\sk}\}$, and the task basis $\taskbasis$ (stages (i) and (ii)).
The \emph{detect} block is the only part paid per suspect: it samples the suspect on $\Xtest$, maps that text through the same probe and projection, and ranks the claimed key (stage (iii)).
The claimed key enters exactly where each decoy does, as an index that selects reference traces, which is what makes the rank test exact.
No step invokes the watermark scheme's own test, so the same algorithm applies unchanged to any keyed watermark.
The computational cost is dominated by the one-time generation of teacher traces under $M+1$ keys (App.~\ref{app:detector-details}).

\section{Experiments and Results}
\label{sec:wild}
Most main-body tables report all three $n$-gram windows of the watermark ($n{=}1,2,3$) at three operating points.
Unless stated otherwise, the text quotes $n$-gram${=}3$, the hardest window for both detectors.

\subsection{Experimental Setup}
\label{subsec:exp-setup}

\myparagraph{Distillation.}
We evaluate distillation detection across $24$ suspect model architectures, $4$ teacher models, and $4$ watermarking schemes.
In the default setting, the teacher is DeepSeek-R1-Distill-Qwen-14B~\citep{guo2025deepseek} watermarked with TextSeal~\citep{sander2026textseal}, a dual-key scheme that samples each token under one of two secret keys, mixed at a fixed rate $\alpha{=}0.1$.
Following standard reasoning-distillation protocols, an adversary fine-tunes a Qwen2.5-3B base model~\citep{qwen2.5} for math on $1{,}900$ quality-filtered OpenR1-Math-220k traces~\citep{openr1} from that teacher.
All traces are generated under one fixed math-teacher system prompt, on which the suspect model is also conditioned during fine-tuning.
We evaluate capability on held-out GSM8K~\citep{cobbe2021training} and Math-500~\citep{hendrycks2021measuring}; App.~\ref{app:distillation-details} gives the full distillation setup.

\myparagraph{Sampling suspect model outputs.}
The verifier chooses one domain to audit and draws every prompt it uses for detection from that domain; it never sees the adversary's prompts or traces (\S\ref{sec:setup}).
By default the audited domain is OpenR1-Math-220k~\citep{openr1}, held out from the adversary's distillation prompts, and every audit stage uses the same system prompt as the distillation traces, which we call a \emph{matched-prompt} audit; \S\ref{subsec:verifier-prompts} varies both the domain and the system prompt.
We read each suspect on $1{,}900$ prompts from the test pool $\Xtest$, disjoint from the distillation prompts $\Xtrain$, and score the first $512$ tokens of every response, around $1$M tokens per suspect.

\myparagraph{Detector.}
\method turns text into a gradient with a probe model, by default a clean Qwen2.5-3B base warmed on around $450$ clean teacher traces from the audited domain (App.~\ref{app:probe-model} studies both the choice of probe and the warm-up).
On $512$ build-pool prompts we query the teacher under the claimed watermark key and $999$ decoys, then map those responses and the suspect's generations through the probe, giving a $1{,}000$-key reference gradient bank and the suspect gradient scored against it.
Before scoring, we project both away from a PCA task subspace fit on $43$ off-the-shelf checkpoints, which leaves the key-specific alignment that the rank test reads (\S\ref{subsec:task-removal}).
All three stages draw separately from the build pool $\Xbuild$, disjoint from $\Xtest$.
Ranking the claimed key gives $p$-values down to $0.001$; unless stated otherwise we report the true positive rate (TPR) at $p{\le}0.01$, \ie at a $1\%$ false positive rate (FPR).
App.~\ref{app:calibration-nulls} validates the false positive rate empirically.

\myparagraph{Baseline.}
We compare with the black-box sampled-token detector of \citet{sander2024watermarking}, to our knowledge the only existing detector for this setting.
It reads the same suspect text as \method, so every comparison is paired.
We run it under its own protocol with both filters that give it power: a position is scored only when its context window occurs in a held-out corpus of $5{,}000$ teacher traces regenerated under the key being scored, and each distinct (context, token) pair is scored once.
Its $p$-value is computed analytically from the score's null distribution rather than by ranking decoys, and we threshold both detectors at the same $p$-value.
App.~\ref{app:detector-details} gives the test used for each watermark scheme and the remaining settings.

\begin{table*}[t]
\centering
\housesetup
\caption{\textbf{Distillation detection across $24$ student architectures.} A single TextSeal-watermarked teacher (DeepSeek-R1-Distill-Qwen-14B) is distilled into $24$ students. 
Each cell is the percentage of up to $50$ distilled models per architecture flagged at a $p$-value of $5\%$, $1\%$, or $0.1\%$; \textbf{Overall} pools all models in that column (1,200 for each), and \texttt{Acc.} is post-distillation GSM8K accuracy. 
}
\label{tab:crossarch-full}
\housetable{%
\begin{tabular}{W}
\toprule
\househead{Student structure}
Qwen3-4B-Instruct              & $88.7$ & \ho{100} & \ho{100} & \ho{100} & \hb{63}  & \hb{26} & \hb{9}  & $89.5$ & \ho{100} & \ho{100} & \ho{100} & \hb{38} & \hb{18} & \hb{8}  & $91.6$ & \ho{100} & \ho{100} & \ho{88} & \hb{8}  & \hb{2}  & \hb{0} \\
Qwen3-4B-Instruct-2507         & $91.4$ & \ho{100} & \ho{100} & \ho{100} & \hb{56}  & \hb{33} & \hb{14} & $91.6$ & \ho{100} & \ho{100} & \ho{100} & \hb{36} & \hb{22} & \hb{8}  & $90.8$ & \ho{98}  & \ho{96}  & \ho{90} & \hb{14} & \hb{4}  & \hb{0} \\
Qwen3-8B-Instruct              & $89.6$ & \ho{100} & \ho{100} & \ho{100} & \hb{49}  & \hb{19} & \hb{2}  & $90.8$ & \ho{100} & \ho{100} & \ho{100} & \hb{20} & \hb{12} & \hb{0}  & $89.9$ & \ho{100} & \ho{100} & \ho{92} & \hb{6}  & \hb{0}  & \hb{0} \\
Qwen2.5-7B-Instruct            & $87.8$ & \ho{100} & \ho{100} & \ho{100} & \hb{70}  & \hb{47} & \hb{12} & $87.9$ & \ho{100} & \ho{100} & \ho{98}  & \hb{42} & \hb{24} & \hb{8}  & $88.8$ & \ho{100} & \ho{92}  & \ho{80} & \hb{18} & \hb{2}  & \hb{0} \\
Qwen2.5-Math-7B-Instruct       & $83.1$ & \ho{100} & \ho{100} & \ho{100} & \hb{42}  & \hb{23} & \hb{2}  & $85.6$ & \ho{100} & \ho{100} & \ho{96}  & \hb{34} & \hb{10} & \hb{2}  & $88.0$ & \ho{100} & \ho{100} & \ho{76} & \hb{12} & \hb{4}  & \hb{0} \\
DeepSeek-R1-Distill-Qwen-7B    & $86.7$ & \ho{100} & \ho{100} & \ho{100} & \hb{44}  & \hb{7}  & \hb{0}  & $87.4$ & \ho{100} & \ho{100} & \ho{98}  & \hb{24} & \hb{4}  & \hb{2}  & $87.0$ & \ho{100} & \ho{96}  & \ho{68} & \hb{0}  & \hb{0}  & \hb{0} \\
Qwen3-8B-Base                  & $88.6$ & \ho{95}  & \ho{91}  & \ho{74}  & \hb{51}  & \hb{28} & \hb{5}  & $89.1$ & \ho{80}  & \ho{70}  & \ho{46}  & \hb{40} & \hb{20} & \hb{8}  & $87.0$ & \ho{62}  & \ho{48}  & \ho{24} & \hb{10} & \hb{6}  & \hb{0} \\
Llama-3.1-8B-Instruct          & $84.1$ & \ho{100} & \ho{100} & \ho{100} & \hb{91}  & \hb{77} & \hb{42} & $84.5$ & \ho{100} & \ho{100} & \ho{88}  & \hb{80} & \hb{52} & \hb{18} & $84.4$ & \ho{98}  & \ho{80}  & \ho{66} & \hb{24} & \hb{4}  & \hb{4} \\
Qwen3-1.7B-Instruct            & $77.6$ & \ho{100} & \ho{100} & \ho{100} & \hb{42}  & \hb{19} & \hb{5}  & $80.1$ & \ho{100} & \ho{100} & \ho{96}  & \hb{28} & \hb{16} & \hb{2}  & $81.7$ & \ho{98}  & \ho{88}  & \ho{70} & \hb{4}  & \hb{0}  & \hb{0} \\
Qwen2.5-3B-Instruct            & $77.7$ & \ho{100} & \ho{100} & \ho{100} & \hb{72}  & \hb{56} & \hb{19} & $77.1$ & \ho{100} & \ho{98}  & \ho{94}  & \hb{50} & \hb{26} & \hb{8}  & $79.7$ & \ho{100} & \ho{90}  & \ho{58} & \hb{18} & \hb{6}  & \hb{0} \\
DeepSeek-R1-Distill-Qwen-1.5B  & $75.7$ & \ho{100} & \ho{100} & \ho{100} & \hb{33}  & \hb{2}  & \hb{2}  & $77.3$ & \ho{100} & \ho{100} & \ho{94}  & \hb{12} & \hb{2}  & \hb{0}  & $74.8$ & \ho{100} & \ho{94}  & \ho{76} & \hb{4}  & \hb{0}  & \hb{0} \\
Llama-Guard-3-8B               & $61.7$ & \ho{100} & \ho{100} & \ho{100} & \hb{100} & \hb{84} & \hb{51} & $62.6$ & \ho{100} & \ho{96}  & \ho{76}  & \hb{82} & \hb{64} & \hb{22} & $63.0$ & \ho{92}  & \ho{70}  & \ho{52} & \hb{24} & \hb{6}  & \hb{2} \\
Qwen3-0.6B-Instruct            & $61.3$ & \ho{100} & \ho{100} & \ho{100} & \hb{72}  & \hb{42} & \hb{16} & $62.3$ & \ho{100} & \ho{98}  & \ho{88}  & \hb{30} & \hb{16} & \hb{2}  & $59.6$ & \ho{96}  & \ho{74}  & \ho{50} & \hb{10} & \hb{0}  & \hb{0} \\
Qwen2.5-1.5B-Base              & $55.9$ & \ho{100} & \ho{100} & \ho{100} & \hb{74}  & \hb{35} & \hb{7}  & $52.2$ & \ho{98}  & \ho{90}  & \ho{66}  & \hb{38} & \hb{20} & \hb{6}  & $58.5$ & \ho{82}  & \ho{62}  & \ho{38} & \hb{12} & \hb{6}  & \hb{0} \\
Qwen2.5-1.5B-Instruct          & $58.2$ & \ho{100} & \ho{100} & \ho{98}  & \hb{74}  & \hb{44} & \hb{16} & $58.5$ & \ho{92}  & \ho{80}  & \ho{54}  & \hb{42} & \hb{10} & \hb{0}  & $54.6$ & \ho{68}  & \ho{52}  & \ho{34} & \hb{8}  & \hb{2}  & \hb{0} \\
Llama-3.2-1B-Instruct          & $36.0$ & \ho{100} & \ho{100} & \ho{98}  & \hb{70}  & \hb{42} & \hb{16} & $35.8$ & \ho{98}  & \ho{92}  & \ho{72}  & \hb{52} & \hb{30} & \hb{16} & $38.7$ & \ho{96}  & \ho{76}  & \ho{44} & \hb{14} & \hb{6}  & \hb{0} \\
Qwen2.5-0.5B-Base              & $31.9$ & \ho{100} & \ho{100} & \ho{91}  & \hb{88}  & \hb{65} & \hb{44} & $30.4$ & \ho{92}  & \ho{78}  & \ho{56}  & \hb{66} & \hb{44} & \hb{12} & $36.8$ & \ho{88}  & \ho{62}  & \ho{20} & \hb{16} & \hb{6}  & \hb{0} \\
Qwen2.5-0.5B-Instruct          & $30.7$ & \ho{100} & \ho{100} & \ho{95}  & \hb{79}  & \hb{54} & \hb{19} & $31.6$ & \ho{94}  & \ho{84}  & \ho{58}  & \hb{54} & \hb{28} & \hb{4}  & $30.4$ & \ho{92}  & \ho{64}  & \ho{26} & \hb{18} & \hb{4}  & \hb{2} \\
SmolLM2-1.7B-Base              & $18.5$ & \ho{95}  & \ho{86}  & \ho{56}  & \hb{26}  & \hb{5}  & \hb{2}  & $18.0$ & \ho{82}  & \ho{58}  & \ho{34}  & \hb{6}  & \hb{0}  & \hb{0}  & $18.0$ & \ho{64}  & \ho{34}  & \ho{12} & \hb{4}  & \hb{0}  & \hb{0} \\
Gemma-2-2B-Base                & $10.8$ & \ho{47}  & \ho{19}  & \ho{7}   & \hb{100} & \hb{98} & \hb{88} & $10.4$ & \ho{12}  & \ho{6}   & \ho{2}   & \hb{86} & \hb{72} & \hb{44} & $5.9$  & \ho{14}  & \ho{4}   & \ho{0}  & \hb{30} & \hb{14} & \hb{0} \\
Llama-Guard-3-1B               & $3.8$  & \ho{100} & \ho{84}  & \ho{56}  & \hb{86}  & \hb{54} & \hb{21} & $4.4$  & \ho{88}  & \ho{64}  & \ho{38}  & \hb{46} & \hb{24} & \hb{10} & $5.8$  & \ho{80}  & \ho{42}  & \ho{16} & \hb{16} & \hb{6}  & \hb{2} \\
Gemma-3-1B-Base                & $1.8$  & \ho{56}  & \ho{26}  & \ho{2}   & \hb{84}  & \hb{44} & \hb{14} & $1.7$  & \ho{38}  & \ho{14}  & \ho{4}   & \hb{46} & \hb{20} & \hb{0}  & $1.8$  & \ho{32}  & \ho{14}  & \ho{4}  & \hb{16} & \hb{4}  & \hb{0} \\
SmolLM2-135M-Instruct          & $1.5$  & \ho{56}  & \ho{28}  & \ho{7}   & \hb{7}   & \hb{5}  & \hb{0}  & $1.6$  & \ho{50}  & \ho{18}  & \ho{10}  & \hb{6}  & \hb{4}  & \hb{0}  & $1.8$  & \ho{34}  & \ho{16}  & \ho{2}  & \hb{4}  & \hb{0}  & \hb{0} \\
SmolLM2-135M-Base              & $1.1$  & \ho{56}  & \ho{26}  & \ho{2}   & \hb{12}  & \hb{5}  & \hb{0}  & $1.4$  & \ho{48}  & \ho{18}  & \ho{6}   & \hb{6}  & \hb{2}  & \hb{0}  & $1.3$  & \ho{32}  & \ho{12}  & \ho{4}  & \hb{0}  & \hb{0}  & \hb{0} \\
\midrule
\textbf{Overall} &  & \hob{92} & \hob{86} & \hob{79} & \hbb{71} & \hbb{52} & \hbb{29} & & \hob{86} & \hob{78} & \hob{66} & \hbb{52} & \hbb{29} & \hbb{12} &  & \hob{80} & \hob{65} & \hob{45} & \hbb{20} & \hbb{7} & \hbb{1} \\
\bottomrule
\end{tabular}%
}
\end{table*}

\subsection{Detection Performance}
\label{subsec:wild-access}

\myparagraph{Effect of the student architecture.}
Does detection depend on which student architecture the adversary selects?
We vary the student across $24$ architectures spanning $6$ families ($135$M to $8.2$B parameters) and distill $50$ models per architecture, for $1{,}200$ distilled models per $n$-gram window, holding the teacher, distillation recipe, probe model, and reference gradient bank fixed.
Pooled over all architectures (the \textbf{Overall} row of Tab.~\ref{tab:crossarch-full}), \method outperforms the baseline at every window and operating point, and at the $5\%$ and $1\%$ thresholds its margin widens with the window.
At $n$-gram${=}3$ and $p{\le}0.001$, the baseline recovers almost nothing ($1.3\%$) while \method detects $45\%$ of distilled models.
Detectability is governed by task acquisition rather than architecture: \method mostly misses students that failed to learn the task, so evading detection means giving up the capability that motivated distillation.

\begin{table}[t]
\centering
\housesetup
\caption{\textbf{Distillation detection across distillation temperatures.} Qwen2.5-3B is distilled from DeepSeek-R1-Distill-Qwen-14B on traces sampled at each temperature, while the detector's reference bank stays fixed at $T{=}1.0$.
Each cell is the percentage of up to $100$ distilled models (one per watermark key) flagged at a $p$-value threshold of $5\%$, $1\%$, or $0.1\%$, \ie the TPR at that false-positive rate.}
\label{tab:access-detection}
\housetablematch{%
\begin{tabular}{U}
\toprule
\househead[U]{Temp.}
$1.0$ & \ho{100} & \ho{100} & \ho{100} & \hb{72} & \hb{43} & \hb{26} & \ho{100} & \ho{100} & \ho{98} & \hb{50} & \hb{25} & \hb{9} & \ho{100} & \ho{99} & \ho{88} & \hb{23} & \hb{9} & \hb{1} \\
$0.7$ & \ho{100} & \ho{100} & \ho{100} & \hb{37} & \hb{16} & \hb{2}  & \ho{100} & \ho{97}  & \ho{88} & \hb{25} & \hb{12} & \hb{0} & \ho{99}  & \ho{93} & \ho{62} & \hb{12} & \hb{2} & \hb{0} \\
$0.5$ & \ho{100} & \ho{100} & \ho{99}  & \hb{24} & \hb{7}  & \hb{2}  & \ho{94}  & \ho{78}  & \ho{40} & \hb{8}  & \hb{2}  & \hb{1} & \ho{71}  & \ho{51} & \ho{22} & \hb{4}  & \hb{2} & \hb{0} \\
\bottomrule
\end{tabular}%
}
\end{table}

\myparagraph{Effect of the distillation temperature.}
What happens when the adversary samples its distillation traces at a temperature the verifier does not know?
We generate the teacher's traces at $T{=}1.0$, $0.7$, and $0.5$ and distill $100$ students from each setting.
A lower $T$ concentrates the teacher's output distribution on its most likely tokens.
For \method, the reference gradient bank remains fixed at $T{=}1.0$, creating a mismatch at the two lower temperatures.
Tab.~\ref{tab:access-detection} shows that lowering the distillation temperature costs the black-box baseline most of its power at every window, whereas \method at $p{\le}0.01$ still flags every suspect at $n$-gram${=}1$ and about half of them at $n$-gram${=}3$, even at $T{=}0.5$.
Temperature thus governs \emph{what the tokens reveal, not what the gradient reveals}: lowering it is a cheap evasion against the baseline but not against \method.

\begin{table}[t]
\centering
\housesetup
\caption{\textbf{Distillation detection across teacher models.} The same Qwen2.5-3B student is distilled at $T{=}1.0$ from each of four teachers spanning three families and $8$B to $120$B parameters, and the detector's reference bank is rebuilt for each teacher.
Cells are read as in Tab.~\ref{tab:access-detection}, at $n$-gram${=}3$, over up to $100$ distilled models per teacher; \texttt{Acc.} is mean student GSM8K accuracy.
The thresholds are nominal $p$-values on the DeepSeek row, as elsewhere, and empirical false-positive rates on the three swapped-teacher rows. Teacher properties are in Tab.~\ref{tab:teacher-props}.}
\label{tab:cross-teacher}
\housetablematch{%
\begin{tabular}{G}
\toprule
\househead[G]{Teacher}
DeepSeek-R1-Distill-Qwen-14B & $68.9$ & \ho{100} & \ho{99}  & \ho{88}  & \hb{23} & \hb{9}  & \hb{1}  \\
gpt-oss-120B    & $77.2$ & \ho{96}  & \ho{76}  & \ho{30}  & \hb{18} & \hb{10} & \hb{2}  \\
QwQ-32B         & $80.8$ & \ho{99}  & \ho{97}  & \ho{91}  & \hb{88} & \hb{74} & \hb{42} \\
Llama-3.1-8B    & $74.9$ & \ho{100} & \ho{100} & \ho{100} & \hb{93} & \hb{87} & \hb{64} \\
\bottomrule
\end{tabular}%
}
\end{table}

\myparagraph{Effect of the teacher model.}
Are some teachers harder to protect than others?
We distill the same Qwen2.5-3B student from each of four teachers spanning three families and $8$B to $120$B parameters: DeepSeek-R1-Distill-Qwen-14B, gpt-oss-120B~\citep{agarwal2025gpt}, QwQ-32B~\citep{qwq32b}, and Llama-3.1-8B~\citep{grattafiori2024llama}, rebuilding the reference bank for each.
Tab.~\ref{tab:cross-teacher} shows that, at $p\leq 0.05$, \method detects $96$ to $100\%$ of distilled models for every teacher.
The baseline loses most of its power on two teachers and approaches \method on the other two.
Its power is thus inherited from the teacher and varies sharply with it, whereas \method is robust across teachers; App.~\ref{app:cross-teacher-details} relates this to how often each student repeats its contexts.

\newcommand{\hnr}{\makebox[\hcw][c]{---}}

\begin{table}[t]
\centering
\housesetup
\caption{\textbf{Distillation detection across watermarking schemes.} The teacher's default TextSeal decoder is replaced by each of three other schemes, with the detector's reference bank rebuilt to match. Cells are read as in Tab.~\ref{tab:access-detection}, over $50$ distilled models per scheme and window. }
\label{tab:watermark-generality}
\housetablematch{%
\begin{tabular}{U}
\toprule
\househead[U]{Watermark}
GumbelMax    & \ho{100} & \ho{100} & \ho{100}  & \hb{72} & \hb{44} & \hb{28} & \ho{100} & \ho{98}  & \ho{98} & \hb{52} & \hb{32} & \hb{6} & \ho{100} & \ho{100} & \ho{100} & \hb{22} & \hb{10} & \hb{0} \\
KGW    & \ho{100} & \ho{100} & \ho{100}  & \hb{58} & \hb{26} & \hb{8}  & \ho{94}  & \ho{86}  & \ho{56} & \hb{40} & \hb{20} & \hb{4} & \ho{90}  & \ho{68}  & \ho{34} & \hb{16} & \hb{2}  & \hb{2} \\
SynthID & \ho{100} & \ho{100} & \ho{100}  & \hb{64} & \hb{38} & \hb{8}  & \ho{100} & \ho{100} & \ho{98} & \hb{44} & \hb{16} & \hb{0} & \ho{100} & \ho{96}  & \ho{78} & \hb{24} & \hb{8}  & \hb{4} \\
\bottomrule
\end{tabular}%
}
\end{table}

\myparagraph{Effect of the watermark scheme.}
Is detection tied to the watermark the teacher happens to deploy?
We replace TextSeal in the teacher's decoder with GumbelMax~\citep{aaronson2022gumbelmax}, KGW ($\gamma{=}0.5$)~\citep{kirchenbauer2023watermark}, and SynthID~\citep{dathathri2024scalable}, rebuilding the reference bank for each and distilling $50$ students per scheme and window.
Tab.~\ref{tab:watermark-generality} shows that all three transfer: \method flags every student at $p{\le}0.01$ and $n$-gram${=}1$ under each scheme, and it outperforms the baseline in every cell, by a margin that widens with the window at $p{\le}0.05$.
\method needs only traces the teacher generated under a key, not the watermark's own detection test.
The baseline, in contrast, scores the suspect's tokens with that test, which must be derived scheme by scheme (App.~\ref{app:detector-details}).
To audit a new keyed watermark, the verifier therefore only generates teacher traces under its keys, rebuilds the reference bank, and runs the same detector.

\begin{table}[ht]
\centering
\housesetup
\caption{\textbf{Distillation detection across the audited domain.} The same $100$ suspects, all distilled on OpenR1-Math~\citep{openr1}, are audited on three math datasets from the distillation domain (top block) and two domains outside it (bottom block), with both detectors built entirely from the audited domain's own prompts and system prompt. 
Cells are read as in Tab.~\ref{tab:crossarch-full}. Detailed setup in App.~\ref{app:ood-domain}.}
\label{tab:ood-domain}

\housetablematch{%
\begin{tabular}{U}
\toprule
\househead[U]{Audited domain}
\emph{OpenR1-Math} & \ho{100} & \ho{100} & \ho{96} & \hb{76} & \hb{49} & \hb{19} & \ho{98} & \ho{92} & \ho{69} & \hb{51} & \hb{24} & \hb{9} & \ho{95} & \ho{83} & \ho{45} & \hb{26} & \hb{12}  & \hb{2} \\
MATH  & \ho{91} & \ho{74} & \ho{56} & \hb{74} & \hb{49} & \hb{20} & \ho{71} & \ho{48} & \ho{15} & \hb{60} & \hb{28} & \hb{7} & \ho{65} & \ho{38} & \ho{9}  & \hb{24} & \hb{7} & \hb{2} \\
GSM8K & \ho{29} & \ho{14} & \ho{3}  & \hb{35} & \hb{13} & \hb{2}  & \ho{18} & \ho{7}  & \ho{1}  & \hb{16} & \hb{9}  & \hb{1}  & \ho{22} & \ho{10} & \ho{1}  & \hb{7}  & \hb{3}  & \hb{0} \\
\midrule
Code  & \ho{12} & \ho{5}  & \ho{0}  & \hb{50} & \hb{19} & \hb{4}  & \ho{5}  & \ho{1}  & \ho{0}  & \hb{31} & \hb{11} & \hb{2}  & \ho{8}  & \ho{4}  & \ho{0}  & \hb{7} & \hb{2}  & \hb{0} \\
Chat  & \ho{11} & \ho{1}  & \ho{0}  & \hb{22} & \hb{4}  & \hb{1}  & \ho{2}  & \ho{0}  & \ho{0}  & \hb{15}  & \hb{4}  & \hb{0}  & \ho{4}  & \ho{1}  & \ho{0}  & \hb{13} & \hb{4}  & \hb{1} \\
\bottomrule
\end{tabular}%
}
\end{table}

\myparagraph{Effect of the audited domain and system prompt.}
\phantomsection\label{subsec:verifier-prompts}
How close must the audited domain be to the adversary's distillation data?
We hold the $100$ suspects fixed and change the domain the verifier audits, from the math domain they were distilled on (OpenR1-Math, MATH, GSM8K) to two domains outside it (code and chat).
For each domain we rebuild the probe warm-up, the task subspace, and both detectors from that domain's own prompts, none of them the teacher's training system prompt; App.~\ref{app:ood-domain} lists the exact prompts.
Tab.~\ref{tab:ood-domain} shows that \method's detection power follows where the suspects were distilled: it ranks every math domain above both non-math ones in every cell, whereas the baseline ranks code above GSM8K at $n$-gram${=}1$ and $2$.
A hit therefore tells the verifier on which domain the suspect's text carries the watermark, not only that it carries one; a low score elsewhere means nothing was found there, not that the model is clean.

\begin{figure*}[t]
  \centering
  \input{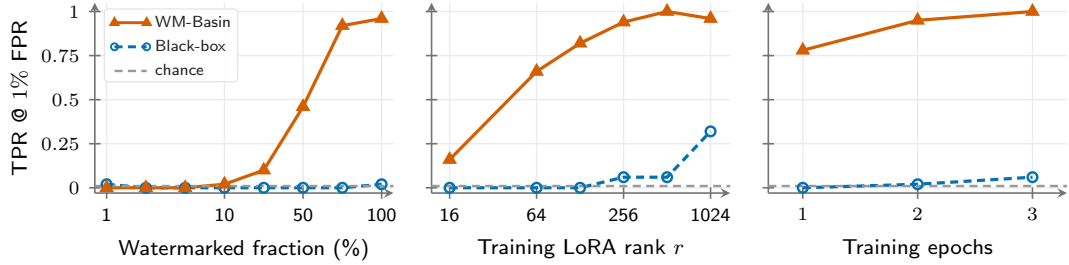}
  \caption{\textbf{What governs detection: each knob swept on its own axis} ($n$-gram${=}3$). Each panel varies \emph{one} knob, with its value on the $x$-axis and the TPR at a fixed $1\%$ FPR on the $y$-axis, for \method and the black-box baseline; the grey dashed line marks chance (TPR${=}$FPR${=}1\%$). Each configuration trains $50$ students, each with a different watermark key. \textbf{Left:} watermarked fraction of the training traces. \textbf{Middle:} LoRA rank $r$. \textbf{Right:} training epochs. The all-clean control (no watermarked traces) is not plotted; none of its $50$ students is flagged at any of the three thresholds.}
  \label{fig:memorization}
\end{figure*}

\myparagraph{Effect of the watermarked fraction and the training recipe.}
What in the adversary's data and training recipe lets the watermark reach the student?
At $n$-gram${=}3$ we sweep one knob at a time around the default recipe: the watermarked fraction of a fixed $1{,}900$-trace training set from $1$ to $100\%$ (the remainder filled with clean traces from the same teacher), the distillation LoRA rank $r$ from $16$ to $1{,}024$, and $1$, $2$, or $3$ training epochs.
Fig.~\ref{fig:memorization} shows that detection power depends mostly on the watermarked fraction and the LoRA rank, and much less on how long the student trains.
(1)~\method performs at chance when at most $10\%$ of the traces are watermarked and exceeds $90\%$ TPR once $75\%$ are.
(2)~Its TPR climbs from $16\%$ at $r{=}16$ to $100\%$ at $r{=}512$.
(3)~A single epoch already yields $78\%$ TPR.
The black-box baseline stays at or below $2\%$ at every fraction and reaches its highest value in the figure at the largest rank; \method outperforms it in every cell of the rank and epoch sweeps.

\subsection{Post-processing and Basin Dynamics}
\label{subsec:persistence}

In \S\ref{subsec:direction-signal}, we showed that a watermarked student sits in a narrow basin: a small step along the watermark direction (Eq.~\eqref{eq:watermark_direction}) leaves it, while a much larger step along the task direction does not.
We now examine in which direction standard post-processing moves the student, and whether it removes the watermark.

\begin{table}[t]
\centering
\housesetup
\caption{\textbf{Distillation detection under compression.} Each row applies one operation to the same $100$ distilled suspects and regenerates their text, holding the probe and the reference bank fixed.
Cells are read as in Tab.~\ref{tab:access-detection}; \texttt{Acc.} is mean GSM8K accuracy per window. Operations below the mid-rule leave an unusable model, with accuracy under $0.9{\times}$ the uncompressed $70.7$.
The black-box cells of the uncompressed row predate the token matching used for the six operations, so compare the black-box columns only across operations.
The weight-space displacement of every operation is in Tab.~\ref{tab:persistence-geometry-full} and the setup in App.~\ref{app:persistence-compression}.}
\label{tab:persistence}
\housetablematch{%
\begin{tabular}{W}
\toprule
\househead[W]{Compression}
none     & $70.7$ & \ho{100} & \ho{100} & \ho{100} & \hb{81} & \hb{50} & \hb{17} & $70.7$ & \ho{100} & \ho{100} & \ho{99} & \hb{44} & \hb{19} & \hb{5} & $70.7$ & \ho{100} & \ho{95} & \ho{80} & \hb{8}  & \hb{4} & \hb{2} \\
int8     & $72.2$ & \ho{100} & \ho{100} & \ho{100} & \hb{75} & \hb{50} & \hb{17} & $71.0$ & \ho{100} & \ho{100} & \ho{99} & \hb{50} & \hb{27} & \hb{6} & $70.0$ & \ho{100} & \ho{99} & \ho{85} & \hb{24} & \hb{7} & \hb{1} \\
prune10  & $77.1$ & \ho{100} & \ho{100} & \ho{100} & \hb{76} & \hb{49} & \hb{17} & $76.2$ & \ho{100} & \ho{100} & \ho{98} & \hb{52} & \hb{27} & \hb{9}  & $77.3$ & \ho{99}  & \ho{94} & \ho{77} & \hb{21} & \hb{8} & \hb{1} \\
\midrule
prune20  & $59.5$ & \ho{100} & \ho{100} & \ho{100} & \hb{70} & \hb{44} & \hb{20} & $61.2$ & \ho{100} & \ho{96} & \ho{82} & \hb{44} & \hb{24} & \hb{10} & $57.6$ & \ho{91} & \ho{82} & \ho{49} & \hb{20} & \hb{4} & \hb{0} \\
int4     & $50.4$ & \ho{79}  & \ho{59}  & \ho{31}  & \hb{50} & \hb{25} & \hb{11}  & $50.8$ & \ho{43}  & \ho{25} & \ho{12} & \hb{38} & \hb{16}  & \hb{2} & $45.4$ & \ho{46} & \ho{25} & \ho{10} & \hb{7}  & \hb{1} & \hb{0} \\
nf4      & $55.8$ & \ho{100} & \ho{100} & \ho{99}  & \hb{70} & \hb{46} & \hb{23}  & $42.9$ & \ho{100} & \ho{95} & \ho{76} & \hb{48} & \hb{26} & \hb{4} & $41.2$ & \ho{95} & \ho{79} & \ho{40} & \hb{8} & \hb{3} & \hb{0} \\
prune30  & $37.8$ & \ho{93}  & \ho{71}  & \ho{36}  & \hb{52} & \hb{30} & \hb{8}  & $39.6$ & \ho{75}  & \ho{51} & \ho{12} & \hb{32} & \hb{13} & \hb{1} & $28.9$ & \ho{49} & \ho{26} & \ho{7}  & \hb{14} & \hb{1} & \hb{0} \\
\bottomrule
\end{tabular}%
}
\end{table}

\myparagraph{Effect of model compression.}
Can an adversary quantize or prune the watermark out of a deployed student?
We apply int8, int4, or nf4 quantization, or $10$ to $30\%$ unstructured pruning, to the same $100$ suspects.
Tab.~\ref{tab:persistence} shows that only int8 and $10\%$ pruning leave a usable model, and both cost \method at most three points at any window and operating point.
The other operations weaken detection but also leave the student unusable.
The weight-space decomposition explains why (Tab.~\ref{tab:persistence-geometry-full}): every operation displaces the student at least three orders of magnitude further orthogonally to the watermark direction than along it.
Compression therefore does not move the student out of the watermark basin, and it remains detectable.

\begin{table}[t]
\centering
\housesetup
\caption{\textbf{Distillation detection under continued training.} Each row continues training the same distilled suspects on clean data, holding the probe, the reference bank, and the generation settings fixed. Cells are read as in Tab.~\ref{tab:access-detection}, one block per $n$-gram window.
$\Delta$\texttt{Acc.} is the paired per-key change in GSM8K accuracy, and \texttt{Along}~$w$ is the signed projection of the weight change on the watermark direction $\dwmhat$ (Eq.~\eqref{eq:watermark_direction}).
\textbf{Open-weight}~\citep{sander2024watermarking} is teacher-forced on the adversary's own distillation traces, an idealized read that our black-box verifier cannot run (App.~\ref{app:continued-sft}).
Cohort sizes and the full weight-space decomposition of every row are in Tab.~\ref{tab:continued-sft-geometry}.}
\label{tab:continued-sft}
\housetablematch{%
\begin{tabular}{K}
\toprule
\househead[K]{Cond.}
\multicolumn{12}{l}{\emph{$n$-gram${=}1$}} \\
SFT (reason)  & $+4.8$ & $-0.252$ & \ho{3}   & \ho{1}   & \ho{0}   & \hb{39} & \hb{15} & \hb{3}  & \hr{92}  & \hr{82}  & \hr{52}  \\
SFT (multi.)  & $+7.0$ & $-0.013$ & \ho{33}  & \ho{16}  & \ho{3}   & \hb{63} & \hb{34} & \hb{13} & \hr{99}  & \hr{97}  & \hr{89}  \\
GRPO (GSM8K)  & $+2.4$ & $0.000$  & \ho{100} & \ho{100} & \ho{100} & \hb{58} & \hb{42} & \hb{25} & \hr{100} & \hr{100} & \hr{92}  \\
GRPO (K\&K)   & $+0.5$ & $0.000$  & \ho{100} & \ho{100} & \ho{100} & \hb{57} & \hb{14} & \hb{0}  & \hr{100} & \hr{100} & \hr{86}  \\
GRPO (IFEval) & $-1.4$ & $0.000$  & \ho{100} & \ho{100} & \ho{100} & \hb{71} & \hb{57} & \hb{14} & \hr{100} & \hr{100} & \hr{86}  \\
\midrule
\multicolumn{12}{l}{\emph{$n$-gram${=}2$}} \\
SFT (reason)  & $+1.0$ & $-0.182$ & \ho{4}   & \ho{2}   & \ho{0}   & \hb{22} & \hb{9}  & \hb{1}  & \hr{99}  & \hr{97}  & \hr{76}  \\
SFT (multi.)  & $+4.6$ & $-0.009$ & \ho{27}  & \ho{13}  & \ho{3}   & \hb{36} & \hb{18} & \hb{5}  & \hr{100} & \hr{100} & \hr{99}  \\
GRPO (GSM8K)  & $+2.2$ & $0.000$  & \ho{100} & \ho{100} & \ho{87}  & \hb{40} & \hb{40} & \hb{7}  & \hr{100} & \hr{100} & \hr{100} \\
GRPO (K\&K)   & $+0.9$ & $0.000$  & \ho{100} & \ho{100} & \ho{100} & \hb{40} & \hb{30} & \hb{20} & \hr{100} & \hr{100} & \hr{100} \\
GRPO (IFEval) & $+1.0$ & $0.000$  & \ho{100} & \ho{100} & \ho{100} & \hb{30} & \hb{30} & \hb{10} & \hr{100} & \hr{100} & \hr{100} \\
\midrule
\multicolumn{12}{l}{\emph{$n$-gram${=}3$}} \\
SFT (reason)  & $+2.4$ & $-0.146$ & \ho{6}   & \ho{1}   & \ho{0}   & \hb{7}  & \hb{3}  & \hb{1}  & \hr{98}  & \hr{95}  & \hr{73}  \\
SFT (multi.)  & $+6.0$ & $-0.007$ & \ho{21}  & \ho{4}   & \ho{0}   & \hb{15} & \hb{5}  & \hb{0}  & \hr{100} & \hr{100} & \hr{97}  \\
GRPO (GSM8K)  & $+0.5$ & $0.000$  & \ho{100} & \ho{93}  & \ho{53}  & \hb{0}  & \hb{0}  & \hb{0}  & \hr{100} & \hr{100} & \hr{100} \\
GRPO (K\&K)   & $-0.2$ & $0.000$  & \ho{100} & \ho{100} & \ho{50}  & \hb{0}  & \hb{0}  & \hb{0}  & \hr{100} & \hr{100} & \hr{100} \\
GRPO (IFEval) & $-0.3$ & $0.000$  & \ho{100} & \ho{100} & \ho{50}  & \hb{10} & \hb{0}  & \hb{0}  & \hr{100} & \hr{100} & \hr{100} \\
\bottomrule
\end{tabular}%
}
\end{table}

\myparagraph{Effect of continued SFT and RL.}
Does continued training on clean data move the distilled model out of its watermark basin?
We continue training the suspects via SFT on cross-domain data and via GRPO~\citep{shao2024deepseekmath} on verifiable tasks (details in App.~\ref{app:continued-sft}), holding the probe and the reference bank fixed.
Tab.~\ref{tab:continued-sft} shows that continued RL leaves the student detectable while continued SFT weakens the signal: \method detects every distilled model in every GRPO arm at $p{\le}0.05$ and every window, whereas both SFT arms substantially reduce its power.
This ordering matches how far each update moves the model along the watermark direction: SFT on general reasoning traces moves it furthest, SFT on multilingual math less, and GRPO barely at all.
The open-weight detector, teacher-forced on the distillation traces, nevertheless still detects the SFT-trained students.
Continued SFT thus weakens the signal observable in the student's outputs without moving the model out of the watermark basin.

\section{Related Work}
\label{sec:related}

\myparagraph{Model extraction and unauthorized distillation.}
Frontier providers, threat-intelligence groups, policy analysts, and security researchers now document distillation-oriented misuse in the wild: adversaries harvest reasoning traces at scale from deployed APIs to train competing models~\citep{anthropic2026distillation,gtig2026distillation,panfilov2026stealing}.
The underlying mechanism is knowledge distillation, which transfers capability from a teacher to a student model~\citep{hinton2015distilling}.
When exposed through an API, it becomes \emph{model extraction}: black-box query access alone, without the target's parameters or training data, suffices to replicate the teacher's functionality~\citep{tramer2016stealing}.
The threat is sharpest for reasoning models, whose step-by-step traces provide dense supervision.
Capability also transfers through synthetic instructions, model-generated reasoning, rubric or reward feedback, and agent trajectories~\citep{wang2023self,guo2025deepseek,gunjal2507rubrics,zeng2024agenttuning}.
We study supervised fine-tuning on collected reasoning traces, motivated by evidence that small, high-quality reasoning sets sharply improve students~\citep{muennighoff2025s1}.

\myparagraph{Defenses against extraction.}
Existing defenses restrict what the API returns, monitor extraction-like querying~\citep{anthropic2026distillation,gtig2026distillation}, or make outputs less useful for training via antidistillation sampling or trace rewriting~\citep{savani2026antidistillation,ma2026protecting}.
Such methods trade away utility for benign users or assume that the attacker stays observable during collection.
That assumption is fragile in practice: adversaries spread collection across many accounts and third-party API resellers, so no single query stream looks like extraction~\citep{anthropic2026distillation}.
Detection is complementary: it acts after the fact and leaves the deployed model untouched.
A parallel black-box line, dataset inference~\citep{maini2021dataset}, resolves ownership by testing whether a suspect shares private knowledge of the victim's data, without inserting a watermark; it observes that inserted watermarks are off-distribution and therefore not always preserved under stealing.
Classical model watermarking instead hides a signal directly in the weights or via backdoors~\citep{uchida2017embedding,adi2018turning}, read from the parameters or by querying a trigger set; both assume the owner watermarked the suspect model under test, whereas our suspect is a model the owner never trained.

\myparagraph{Watermarking and radioactivity.}
Text watermarking embeds a secret statistical signal into generated text~\citep{kirchenbauer2023watermark,kuditipudi2023robust,sander2026textseal}.
The idea of tracing data through a trained model originates in computer vision: \citet{sablayrolles2020radioactive} show that imperceptible ``radioactive'' watermarks in a dataset leave a statistical residue in models trained on it, detectable with a calibrated $p$-value even when only $1\%$ of the data is marked.
Our rank test follows the same framework.
\citet{sander2024watermarking} carry radioactivity over to language models, showing that watermarked distillation leaves a key-specific residue in the student, and others study how well such watermarks survive distillation~\citep{sander2026textseal,zhao2022distillation,pan2025can}.
The black-box verifier of \citet{sander2024watermarking}, which counts watermark-favored tokens in a suspect's sampled text, defines the setting we work in and is, to the best of our knowledge, the only existing method for the task; it is our baseline throughout.
Its limits were not previously characterized, and establishing them is part of our contribution.
The statistic accumulates evidence only where exact watermarked contexts recur, which we show is fragile to wider windows, lower decoding temperatures, and larger text budgets (\S\ref{subsec:wild-access}, App.~\ref{app:token-budget}), the same failure mode that paraphrase attacks exploit against surface-level watermark detection~\citep{sadasivan2023can}.
\method keeps the access level and the generations, and replaces the token statistic with a weight-space one.

\myparagraph{Weight-space geometry and basins.}
\method relies on the fact that students distilled under one key settle into a shared, key-specific region of weight space.
This connects to the geometry of neural loss landscapes: independently trained optima are joined by simple, near-constant-loss curves~\citep{garipov2018loss}, and once permutation symmetries are accounted for, the landscape resembles a single wide \emph{basin}, so separately trained models can be aligned by permutation and merged in weight space~\citep{ainsworth2022git}.
Large language models exhibit the same basin-like structure~\citep{chen2026unveiling,li2018visualizing,peng2024navigating}.
\method treats the watermark key as carving out one such basin and asks whether a suspect has fallen into it.

\myparagraph{Gradient attribution and representation post-processing.}
To read this geometric signal from text alone, \method backpropagates a suspect's completions through a fixed probe model.
This draws on first-order data attribution: influence functions trace a prediction back to the training points responsible for it using only gradient access~\citep{koh2017understanding}, an idea scaled by gradient tracing~\citep{pruthi2020estimating} and random-projected attribution~\citep{park2023trak}, and adapted to instruction tuning with low-rank projected gradients~\citep{xia2024less}.
Intuitively, a training example's gradient captures how it would nudge the model, so aligned gradients indicate similar examples.
Raw gradient cosines, however, are dominated by a few shared directions (generic language and formatting): a handful of top components inflate the similarity between otherwise unrelated items~\citep{ethayarajh2019contextual,gao2019representation,radovanovic2010hubs,dinu2014improving}.
We remove these with an ``All-but-the-Top'' projection~\citep{mu2017all}, stripping the few dominant yet uninformative directions so that the faint key-specific signal stands out from the generic task geometry.

\section{Conclusion and Future Work}
\label{sec:conclusion}
In this work, we introduce the \emph{watermark basin}: distilling on watermarked traces encodes the teacher's secret key as a faint direction in parameter space, displacing the student's weights along that key direction and into a region that later training does not easily escape.
Building on this geometry, we design \method, which reformulates black-box detection of unauthorized distillation from a fragile, surface-level token count into a geometric alignment test in parameter space.
Across a wide range of evaluations, spanning six student families, four teachers, and four watermarking schemes, \method outperforms the existing baseline by a large margin, and it survives quantization, pruning, and RL post-training.

\myparagraph{Scope and outlook.}
A watermark can be present in a distilled model but unreadable in its text: continued SFT on clean data suppresses the text-level signal while an open-weight probe still recovers the residual, and a supplementary study (App.~\ref{app:paraphrase}) finds the same for rewriting the teacher's traces before distillation.
\method is also an in-domain test: a hit tells the verifier on which domain the suspect's text carries the watermark, not only that it carries one, and a low score elsewhere means nothing was found there, not that the model is clean (\S\ref{subsec:verifier-prompts}).
Both are limits on what a black-box verifier can observe, not on what distillation transfers.
Narrowing the gap between the black-box read and the open-weight probe is thus a promising direction for future work.

\subsection*{Ethics Statement}
This work studies whether a model provider can establish, from API access alone, that a suspect model was distilled from its watermarked teacher.
It involves no human subjects and no personal or user data: the distillation corpus (OpenR1-Math-220k), the utility benchmarks (GSM8K and Math-500), and every teacher, student, probe, and clean-pool checkpoint are publicly released artefacts used under their own licenses, and every experiment runs on models we host ourselves.
Because the method produces evidence of provenance, we report how it behaves on models that are not guilty rather than a bare score.
We also measure what weakens the audit, including rewriting the teacher's traces before distillation, continued fine-tuning, and probing away from the distillation domain.
These are variants of attacks already described in the text-watermarking literature, and we report them because a verifier that does not know the limits of its own test is the more likely harm.

\subsection*{Reproducibility Statement}
The configuration behind every number in the paper is recorded in App.~\ref{app:setup}.
Tab.~\ref{tab:config-defaults} gives the default models, watermark and sampling settings, distillation data and LoRA hyperparameters, probe warm-up, and detector and baseline settings, down to the key ranges and seeds.
Tab.~\ref{tab:config-experiments} indexes every study, the factor it varies, the cohort and sample size, and where its result is reported, so a single arm is reproduced by reading the defaults plus the one row that changes them.
The detector is specified procedurally in Alg.~\ref{alg:wm-basin}, and App.~\ref{app:distillation-details} documents trace generation and the quality filter that takes $5{,}000$ sampled traces to the $1{,}900$ used for training.

\clearpage
\bibliographystyle{ieeenat_fullname}
\bibliography{references}

\clearpage
\beginappendix
\startcontents[appendix]
\section*{Appendix Outline}
\printcontents[appendix]{}{1}{\setcounter{tocdepth}{2}}

\section{The Key-Randomization Rank Test}
\label{app:key-cal}
Token-level detectors read significance from a parametric null for their score: a Gaussian or binomial tail for a green-list count~\citep{kirchenbauer2023watermark,sander2024watermarking}, or a Gamma tail for a Gumbel score~\citep{fernandez2023three}, which is the form our black-box baseline uses (App.~\ref{app:detector-details}).
\method assumes no such model for $\basinscore$ and reads significance from the rank of the claimed key among $M$ decoy keys drawn from the same generator~\citep{dwass1957modified,phipson2016permutation}.

Let $\Hnull$ be the null hypothesis of \S\ref{sec:setup}: the suspect's outputs on the audited domain carry no information about the claimed key $\sk^\star$, as holds when the suspect was never trained within that key's basin.
To test it, we fix the suspect model, the prompts, and the detection pipeline, then draw $M$ independent decoy keys $\sk_1, \dots, \sk_M \leftarrow \mathsf{KeyGen}$.
We evaluate $\basinscore$ at $\sk^\star$ and at each decoy, where a higher score means stronger basin alignment, and form the empirical $p$-value of Eq.~\eqref{eq:rank-pvalue}, which measures how uniquely the suspect aligns with $\sk^\star$ relative to random basins.
The leading $1$ counts $\sk^\star$ among the $M{+}1$ candidates, which prevents $\prank{=}0$ and keeps Type-I error control valid under finite Monte Carlo sampling~\citep{phipson2016permutation}.

The $p$-value is valid because nothing in the pipeline can favor the claimed key under $\Hnull$.
The claimed key and the $M$ decoys are drawn the same way, and the scoring rule uses a key only as an index that selects reference traces; the probe, the random projection, the build prompts, and the fitted $\taskbasis$ are identical for all $M{+}1$ candidates.
The claimed key is therefore equally likely to occupy any of the $M{+}1$ ranks, assuming no ties.
Thresholding its rank $p$-value then controls the FPR in finite samples: the FPR is at most $\alpha$ for any threshold, and equals $\alpha$ when $\alpha$ lies on the attainable rank grid.
The $\ge$ in Eq.~\eqref{eq:rank-pvalue} counts ties against rejection, so ties cost power rather than validity.
None of this needs a parametric model of $\basinscore$~\citep{pesarin2010permutation}: Eq.~\eqref{eq:rank-pvalue} places no restriction on the score, so the guarantee survives arbitrary non-linear transformations of the suspect model.
App.~\ref{app:calibration-nulls} measures the realized rate against the nominal one for every deployed key.

The smallest attainable $p$-value is $1/(M{+}1)$, so the decoy count sets the finest significance an audit can certify: reporting at $\alpha{=}0.001$ requires $M \ge 999$ decoys.
Decoy scoring is offline and reuses the build-once artefacts, so improving the resolution can be cheap in practice.

\FloatBarrier

\section{Experimental Setup}
\label{app:setup}
This section presents the experiment setup in \S\ref{sec:wild}.

\subsection{Configuration Summary}
\label{app:config}
Table~\ref{tab:config-defaults} gives in full the default models and numeric settings that \S\ref{subsec:exp-setup} states in summary, and Tab.~\ref{tab:config-experiments} indexes the studies built on them, what each varies and where it is reported.
{\footnotesize
\setlength{\tabcolsep}{4pt}
\renewcommand{\arraystretch}{0.88}
\setlength{\LTcapwidth}{\textwidth}
\setlength{\LTpre}{\medskipamount}\setlength{\LTpost}{\medskipamount}
\begin{longtable}{@{}l p{0.75\linewidth}@{}}
\caption{\textbf{Default experimental configuration.} Models, watermark, distillation, probe, and detector settings used throughout \S\ref{sec:wild} unless stated otherwise; see App.~\ref{app:distillation-details} for full details. The TextSeal parameter $\alpha{=}0.1$ is held fixed in every experiment.}
\label{tab:config-defaults}\\
\toprule
Component & Setting \\
\midrule
\endfirsthead
\multicolumn{2}{@{}l}{\emph{Table~\ref{tab:config-defaults}, continued.}}\\
\toprule
Component & Setting \\
\midrule
\endhead
\bottomrule
\endlastfoot
\multicolumn{2}{@{}l}{\textit{Models}} \\*
Teacher (default) & DeepSeek-R1-Distill-Qwen-14B \\
Teacher (swaps) & gpt-oss-120B, QwQ-32B, Llama-3.1-8B \\
Student (default) & Qwen2.5-3B \\
Student (cross-arch.) & $24$ structures, $135$M to $8.2$B (Llama(-Guard), Qwen2.5/3, QwQ, DeepSeek-Distill, Gemma, SmolLM2) \\
Probe model & Qwen2.5-3B (base, warmed) \\
Clean PCA pool & $43$ off-the-shelf checkpoints ($15$ families, $135$M to $120$B; $129$ cells total; Tab.~\ref{tab:clean-pool}) \\
\midrule
\multicolumn{2}{@{}l}{\textit{Watermark \& teacher sampling}} \\*
Scheme & TextSeal, $\alpha{=}0.1$ \\
Sampling & vLLM; $T{=}1.0$, top-$p{=}0.95$; ${\le}8{,}192$ tokens \\
Distillation temperatures & $\{1.0, 0.7, 0.5\}$ (reference fixed at $T{=}1.0$) \\
\midrule
\multicolumn{2}{@{}l}{\textit{Distillation (student)}} \\*
Dataset & OpenR1-Math-220k \\
System prompt & one fixed math-teacher prompt for every trace; by default every verifier stage (warm-up, bank, clean pool, audit) renders this same prompt, so the default audit is prompt-matched. Only the domain sweep replaces it (App.~\ref{app:ood-domain}) \\
Traces & $5{,}000$ sampled $\to$ quality-filtered $\to$ $1{,}900$ used for training \\
Trace filter & closing \texttt{</think>}, \verb|\boxed{}|, no ${\ge}3{\times}$ repetition, \texttt{math\_verify} match \\
Fine-tuning & LoRA $r{=}512$, lr $2\mathrm{e}{-5}$, $3$ epochs, effective batch $16$, max sequence length $8192$ \\
LoRA targets & $\{q,k,v,o,\mathrm{gate},\mathrm{up},\mathrm{down}\}_{\mathrm{proj}}$ \\
Eval benchmarks & GSM8K, Math-500 \\
\midrule
\multicolumn{2}{@{}l}{\textit{Probe warm-up}} \\*
Data & $450$ clean teacher traces, on build-pool prompts $\Xbuild$ from the domain being audited (App.~\ref{app:ood-domain}) \\
Adapter & LoRA $r{=}128$, $4$ epochs, lr $2\mathrm{e}{-4}$, batch $1$ \\
\midrule
\multicolumn{2}{@{}l}{\textit{Detector (\method)}} \\*
Reference bank & $100$ deployed keys, one student distilled per key; each ranked against a $1{,}000$-key bank ($\sk^\star$ ${+}$ $999$ decoys) \\
Per-key gradient & average of $512$ traces (${\ge}512$ tokens) on build-pool prompts $\Xbuild$ \\
Projection & fixed random projection to $8{,}192$-dim (TRAK) \\
Task subspace & top-$q{=}100$ PCA (all-but-the-top), fit on clean pool \\
$p$-value floor & $0.001$ \\
Operating point & threshold $p{\le}0.01$ (TPR@$1\%$~FPR) \\
\midrule
\multicolumn{2}{@{}l}{\textit{Baseline}~\citep{sander2024watermarking}} \\*
Black-box $n$-gram & corpus $\varphi$ from keyed teacher traces on $5{,}000$ held-out prompts \\
Windows & $n$-gram $\in\{1,2,3\}$ \\
Deduplication & each distinct (context, token) pair scored once across traces \\
Test & baseline's own analytic $p$-value: two-key score fused at the generation $\alpha{=}0.1$, variance-corrected Gamma tail (App.~\ref{app:detector-details}); plain, not entropy-weighted \\
\midrule
\multicolumn{2}{@{}l}{\textit{Suspect text (shared by every detector)}} \\*
Budget & $1{,}900$ held-out prompts $\to$ median $1{,}875$ scored responses, ${\approx}10$M sampled tokens of which ${\approx}1$M are scored (App.~\ref{app:prompt-pools}) \\
Prompts & test pool $\Xtest$, disjoint from $\Xtrain$ and $\Xbuild$ (App.~\ref{app:prompt-pools}) \\
System prompt & the distillation prompt above, i.e.\ matched; replaced only in App.~\ref{app:ood-domain} \\
Sampling & $T{=}1.0$, top-$p{=}0.95$, ${\le}8{,}192$ new tokens, repetition penalty $1.0$, seed $42$ \\
\end{longtable}
}

{\footnotesize
\setlength{\tabcolsep}{4pt}
\renewcommand{\arraystretch}{1.05}
\setlength{\LTcapwidth}{\textwidth}
\setlength{\LTpre}{\medskipamount}\setlength{\LTpost}{\medskipamount}
\begin{longtable}{@{}p{0.18\linewidth} p{0.34\linewidth} p{0.27\linewidth} p{0.15\linewidth}@{}}
\caption{\textbf{Per-experiment settings.} What each study varies, the cohort used, and where the result is reported; varied factors are read against the defaults of Tab.~\ref{tab:config-defaults}. 
}
\label{tab:config-experiments}\\
\toprule
Experiment & Varied factor and levels & Model cohort / sample size & Result \\
\midrule
\endfirsthead
\multicolumn{4}{@{}l}{\emph{Table~\ref{tab:config-experiments}, continued.}}\\
\toprule
Experiment & Varied factor and levels & Model cohort / sample size & Result \\
\midrule
\endhead
\bottomrule
\endlastfoot
Student architecture & $24$ structures, $135$M to $8.2$B parameters & $50$ students per architecture ($1{,}200$ total) & Tab.~\ref{tab:crossarch-full} \\
Distillation temperature & Teacher sampling $T \in \{1.0, 0.7, 0.5\}$; reference bank fixed at $T{=}1.0$ & $100$ students per temperature & Tab.~\ref{tab:access-detection} \\
Teacher model & $4$ teachers spanning $3$ families, $8$B to $120$B; bank rebuilt per teacher; the three swapped teachers distil for $8$ epochs, not $3$ &$100$ students per teacher & Tabs.~\ref{tab:cross-teacher}, \ref{tab:teacher-props} \\
Watermark scheme & $4$ watermark schemes at $n$-gram $\in \{1,2,3\}$ & $50$ students per scheme and window & Tabs.~\ref{tab:watermark-generality}, \ref{tab:watermark-generality-entropy} \\
Audited domain & $5$ domains; every detector stage and the system prompt rebuilt per domain
& Fixed pool of $100$ students & Tab.~\ref{tab:ood-domain} \\
Model compression & int8, int4 and nf4 quantization; $10$ to $30\%$ pruning & $100$ students per operation & Tabs.~\ref{tab:persistence}, \ref{tab:persistence-geometry-full}; Fig.~\ref{fig:postproc-standard} \\
Continued SFT and RL & Post-distillation training: SFT domain, or GRPO task & $100$ students (SFT); $10$ to $20$ (GRPO) & Tabs.~\ref{tab:continued-sft}, \ref{tab:continued-sft-geometry}, \ref{tab:crossdomain-rl} \\
Probe model & $10$ probe checkpoints; reference bank rebuilt per probe & Fixed pool of $100$ students & Tab.~\ref{tab:probe-model} \\
Probe warm-up & Presence, corpus domain, LoRA rank, learning rate, epochs; one knob at a time, $n$-gram${=}3$ only & Fixed pool of $100$ scored keys, reused across all $15$ arms & Tab.~\ref{tab:warmup-ablation} \\
Trace paraphrasing & Same-family Qwen paraphraser at $3$ strengths & $20$ students per strength & Tab.~\ref{tab:paraphrase} \\
Watermarked data fraction & Proportion $f$ of watermarked traces, at a fixed $1{,}900$-trace training budget & $50$ students per fraction & Fig.~\ref{fig:memorization} \\
Student training recipe & LoRA rank $r$, epochs & $50$ students per configuration & Fig.~\ref{fig:memorization} \\
Distillation utility & Clean vs.\ watermarked teacher traces & $100$ watermarked students per $n$-gram window, plus a clean control & Fig.~\ref{fig:distillation-accuracy} \\
Operating-point ROC & Detection threshold, at $n$-gram $\in \{1,2,3\}$ & $100$ watermarked students against $6{,}500$ null cases & Fig.~\ref{fig:roc-by-ngram} \\
Suspect-text budget & Cumulative prefixes of $5$ to $1{,}878$ generations ($0.03$ to $10.3$M tokens supplied) & Fixed pool of $100$ students & Fig.~\ref{fig:text-budget} \\
\midrule
Per-key false-positive rate & Which bank key is read as the deployed one, at $\alpha{=}5\%$ and $n$-gram $\in \{1,2,3\}$ & All $1{,}000$ bank keys, each across $129$ held-out clean model-temperature cells & Tab.~\ref{tab:perkey-fpr} \\
Watermark basin geometry & Weight-space offset along $\dwmhat$ and $\dtaskhat$ & One student distilled under a single key & Fig.~\ref{fig:anisotropic-basin} \\
Task-subspace removal & Removed components $q$: seven settings, from \emph{no PCA} through \emph{mean only} to $q{=}100$ & $1{,}200$ distilled students at all $n$-grams, against $129$ clean model-temperature cells & Fig.~\ref{fig:dehub-operating-point} \\
Rank $p$-value calibration & Significance threshold $\alpha$, swept under the null & $100$ deployed keys across $65$ held-out clean model-temperature cells & Fig.~\ref{fig:calib-curve-perkey} \\
\end{longtable}
}

\subsection{Distillation and Student Training}
\label{app:distillation-details}
The student follows the fine-tuning recipe of Tab.~\ref{tab:config-defaults}. A teacher trace is retained only if it contains a closing \texttt{</think>} tag and a \verb|\boxed{...}| answer, has no substring repeated ${\ge}3$ times under the sliding $100$-character repetition check, and its extracted answer matches the gold answer under \texttt{math\_verify}. The probe $\probe$ is warmed up on clean teacher traces with LoRA.

\begin{figure}[h]
  \centering
  \input{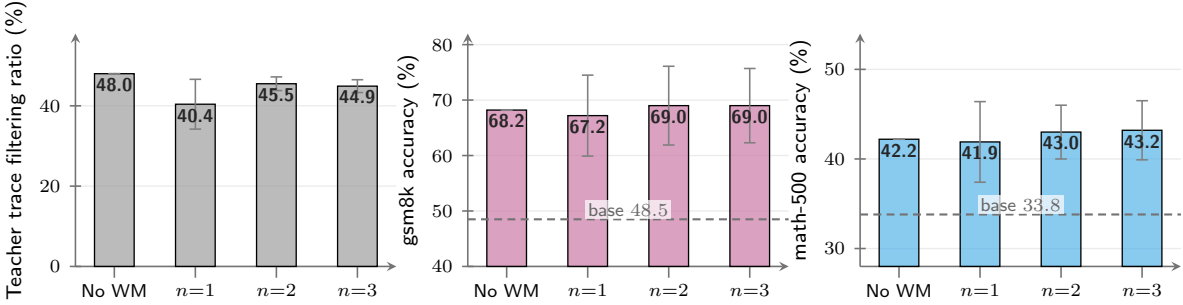}
  \caption{\textbf{Watermarking preserves the utility gained through distillation.} DeepSeek-R1-Distill-Qwen-14B~\citep{guo2025deepseek} is distilled into Qwen2.5-3B~\citep{qwen2.5} using clean traces or TextSeal traces at three $n$-gram windows. The left panel reports the fraction of $5{,}000$ traces passing the quality filter; the other panels report GSM8K and Math-500 accuracy. Watermarked bars average $100$ keyed students with $\pm1$ standard-deviation error bars, while No WM is one student trained on clean traces; dashed lines mark the un-distilled base.
  }
  \label{fig:distillation-accuracy}
\end{figure}

\paragraph{Distillation performance.}
Watermarking the teacher does not cost the student the utility distillation buys. Figure~\ref{fig:distillation-accuracy} compares clean and TextSeal distillation under this recipe on GSM8K~\citep{cobbe2021training} and Math-500~\citep{hendrycks2021measuring}: distillation lifts both benchmarks well above the un-distilled base, and the watermarked students match the clean-distilled control at trace-filter pass rates within $8$ points of it.

\FloatBarrier
\subsection{Prompt Pools and Suspect Sampling}
\label{app:prompt-pools}
Section~\ref{sec:setup} splits the audited domain into disjoint training, build and test pools. In the following, we explain how we split the prompts in our evaluation.
\begin{itemize}
	\item  The \emph{training} pool $\Xtrain$ is the adversary's: OpenR1-Math-220k prompts sampled under the fixed math-teacher system prompt, yielding $5{,}000$ traces, of which $40$ to $48\%$ pass the filter of App.~\ref{app:distillation-details} (Fig.~\ref{fig:distillation-accuracy}) and $1{,}900$ become the distillation set. 
	\item The \emph{build} pool $\Xbuild$ is the verifier's offline side and supplies the three artefacts \method is made of: the ${\approx}450$ clean teacher traces that warm the probe, the $512$ keyed traces averaged into each of the $1{,}000$ reference directions, and the free generations of the $43$ clean checkpoints that fit the task subspace. The baseline's $n$-gram corpus $\varphi$ is a fourth offline artefact, drawn by construction from held-out teacher traces the suspect is never read on. 
	\item  The \emph{test} pool $\Xtest$ is $1{,}900$ held-out prompts and is the only pool a verdict is ever read on. No prompt the adversary trained on is used to build either detector, and no prompt used to build either detector is used to score a suspect. 

\end{itemize}

\FloatBarrier
\subsection{Detector and Baseline Implementation}
\label{app:detector-details}
This section gives the implementation behind the detector and baseline of \S\ref{subsec:exp-setup}.

\myparagraph{Reference key gradient bank and projection.}
For each candidate key, we generate teacher traces on the build pool $\Xbuild$ (App.~\ref{app:prompt-pools}), discard responses below the minimum token length in Tab.~\ref{tab:config-defaults}, and average the surviving reference gradients into $\dwm{\sk}$. A single fixed random projection, shared by all keys and applied unchanged to the suspect gradient $\dsusp$ so that every cosine the rank test compares is taken in one coordinate space, approximately preserves their inner products~\citep{johnson1984extensions,xia2024less}. We compute this with TRAK~\citep{park2023trak}. The deployed key and i.i.d.\ decoys form the bank used by the rank test in Eq.~\eqref{eq:rank-pvalue}, and its size sets the reported $p$-value floor.

\myparagraph{Compute cost.}
Nearly all of the cost is one-time and suspect-independent. Warming the probe is one short LoRA pass (App.~\ref{app:probe-model}). Building the bank dominates: each projected direction averages $512$ reference traces at about $27$ minutes of single-GPU time, so around $450$ GPU-hours for the full $1{,}000$-key bank, which occupies $31.2$\,MiB once projected. Both artefacts are reused by every subsequent audit and are rebuilt only when what they describe changes, namely the teacher or the probe for the bank, and the probe or the prompt distribution for the task subspace (App.~\ref{app:calib-task-removal}). Auditing one suspect then costs sampling its text, the ${\approx}1$M sampled tokens of App.~\ref{app:prompt-pools}, a query cost identical for every detector, plus one backward pass and projection per completion; at the throughput above, the median $1{,}875$-completion budget is roughly $1.6$ GPU-hours on one device. 

\myparagraph{Sampled-token baselines.}
The baselines of~\citet{sander2024watermarking} score a single token per position. In the open-weight setting, teacher-forced reading evaluates the student's top-1 prediction on teacher-generated test text. Their black-box variant scores tokens sampled by the student. We score a token only if its preceding $n$-gram appears in held-out teacher text generated with the key being tested. We exclude tokens whose preceding $n$-gram also appears in the prompt. Following~\citet{sander2024watermarking}, both variants score distinct (context, token) pairs across traces. The open-weight variant additionally requires distinct contexts within each trace.

Seeing only text, the verifier re-tokenizes each suspect response with the \emph{teacher's} tokenizer and scores every position of that stream: the context $n$-gram $c_i$ and the scored token $y_i$ both come from the suspect's own tokens, with no teacher forcing and no student logits. TextSeal's secret is a key pair $(\sk_a,\sk_b)$, and the generator draws its Gumbel noise from $\sk_a$ with probability $\alpha$ and from $\sk_b$ otherwise, so no token reveals which key drove it. We score both and weight them by that generation prior,
\begin{equation}
S_i \;=\; \alpha\big({-}\log(1-r^{(a)}_i)\big) \;+\; (1-\alpha)\big({-}\log(1-r^{(b)}_i)\big),
\qquad
r^{(\cdot)}_i \;=\; \mathrm{PRF}\big(c_i,\, y_i;\, \sk_{\cdot}\big),
\end{equation}
at the deployed $\alpha{=}0.1$.
Here $r_i$ is the keyed pseudorandom value of the scored token: a watermarked teacher prefers tokens with large $r_i$, so a radioactive student produces positions where $S_i$ is larger than chance.
Under $\Hnull$ each ${-}\log(1-r)\sim\mathrm{Exp}(1)$ and the two keys are independent, so $\mathbb{E}[S_i]{=}1$ but $\mathrm{Var}[S_i]{=}\sigma^2_\alpha\equiv\alpha^2{+}(1{-}\alpha)^2{=}0.82$.
Matching $S=\sum_{i=1}^{L}S_i$ over the $L$ retained positions to a Gamma of mean $L$ and variance $L\sigma^2_\alpha$ (shape $L/\sigma^2_\alpha$, scale $\sigma^2_\alpha$) gives
\begin{equation}
\label{eq:bb-gamma}
p \;=\; Q\!\left(\frac{L}{\sigma^2_\alpha},\; \frac{S}{\sigma^2_\alpha}\right),
\qquad
z \;=\; \frac{S-L}{\sigma_\alpha\sqrt{L}},
\end{equation}
with $Q$ the upper regularized incomplete gamma function.
The $p$-value is the probability that a clean model accumulates a total score of at least $S$ over $L$ positions, and $z$ is the same evidence expressed in standard deviations above the null mean.

The three other watermark schemes in App.~\ref{app:watermark-generality} are read on the same principle: GumbelMax is the single-key case, where $\alpha{=}1$ leaves $\sigma^2_\alpha{=}1$,
whereas KGW and SynthID score each position with a keyed-vocabulary indicator and a tournament $g$-value rather than a Gumbel variate.

\FloatBarrier
\subsection{Removing Shared Task Directions}
\label{app:calib-task-removal}
Section~\ref{subsec:task-removal} projects the shared task subspace out of every gradient before scoring; this section gives how that subspace is fit and how many components we remove.

The idea of removing the shared task direction is aligned with removing the shared common vector for the word embeddings studied in~\cite{mu2017all}. Following their postprocessing, we fit PCA to gradients from the clean checkpoint pool and keep its top $q$ principal directions as the shared task basis $\taskbasis$, then project $\dsusp$ and every $\dwm{\sk}$ onto the orthogonal complement of $\taskbasis$ before computing the cosine similarity score $\basinscore(\sk)$ in stage (iii). The number of removed directions, $q$, is the one free parameter of this procedure.

\begin{table}[h]
\centering
\scriptsize
\setlength{\tabcolsep}{4pt}
\renewcommand{\arraystretch}{1.0}
\caption{\textbf{The clean PCA pool.} The $43$ off-the-shelf checkpoints whose gradients fit the shared task subspace of App.~\ref{app:calib-task-removal}, grouped by pretraining lineage. Each model generates on $1{,}000$ build-pool prompts at each of three decoding temperatures ($0.7$, $1.0$, $1.3$).}
\label{tab:clean-pool}
\begin{tabular}{@{}l p{0.76\linewidth} r@{}}
\toprule
Family & Checkpoints & Ckpts. \\
\midrule
Qwen & Qwen2.5-0.5B, -0.5B-Instruct, -1.5B, -1.5B-Instruct, -3B, -3B-Instruct, -7B-Instruct, -Math-7B; Qwen3-0.6B, -1.7B, -4B, -4B-Instruct, -8B, -8B-Base, -32B; QwQ-32B; Qwen1.5-MoE-A2.7B & $17$ \\
Llama & Llama-2-7B, Llama-3.1-8B-Instruct, Llama-3.2-1B-Instruct, Llama-Guard-3-1B, Llama-Guard-3-8B & $5$ \\
DeepSeek & DeepSeek-R1-Distill-Qwen-1.5B, -7B, -14B$^\dagger$; DeepSeek-MoE-16B-Base & $4$ \\
SmolLM2 & SmolLM2-135M, -135M-Instruct, -1.7B & $3$ \\
Gemma & Gemma-2-2B, Gemma-3-1B-PT & $2$ \\
Phi & Phi-3.5-mini-Instruct, Phi-4 & $2$ \\
OLMo & OLMo-2-1B, OLMoE-1B-7B & $2$ \\
Mistral & Mistral-7B-Instruct & $1$ \\
Yi & Yi-1.5-9B-Chat & $1$ \\
Falcon & Falcon3-7B-Instruct & $1$ \\
Granite & Granite-3.1-8B-Instruct & $1$ \\
StarCoder & StarCoder2-7B & $1$ \\
Nemotron & Nemotron-Mini-4B & $1$ \\
Moonlight & Moonlight-16B-A3B & $1$ \\
gpt-oss & gpt-oss-120B & $1$ \\
\midrule
$15$ families & & $43$ \\
\bottomrule
\end{tabular}
\end{table}

\myparagraph{What the clean pool is.}
Table~\ref{tab:clean-pool} lists all models used for removing the task direction: $43$ checkpoints are off-the-shelf, never-distilled causal LMs spanning $15$ pretraining families and $135$M to $120$B parameters. 
None has been fine-tuned on the audited domain and none has ever been exposed to a watermark key. Every checkpoint contributes one cell per decoding temperature ($0.7$, $1.0$, $1.3$), giving the $129$ cells of Tab.~\ref{tab:config-defaults}. 
A cell free-generates on the build pool $\Xbuild$, and its gradients are taken at the same probe model $\probe$, so a clean cell and a suspect differ only in the text they produced. For fixed-threshold TPR evaluation, clean models are not evaluation models, so the PCA fit uses all $129$ cells. The ROC curves and the per-key calibration curves use the balanced subset, i.e., $43$ checkpoints at three temperatures split into disjoint fit and evaluation halves.

\begin{figure}[h]
  \centering
  \input{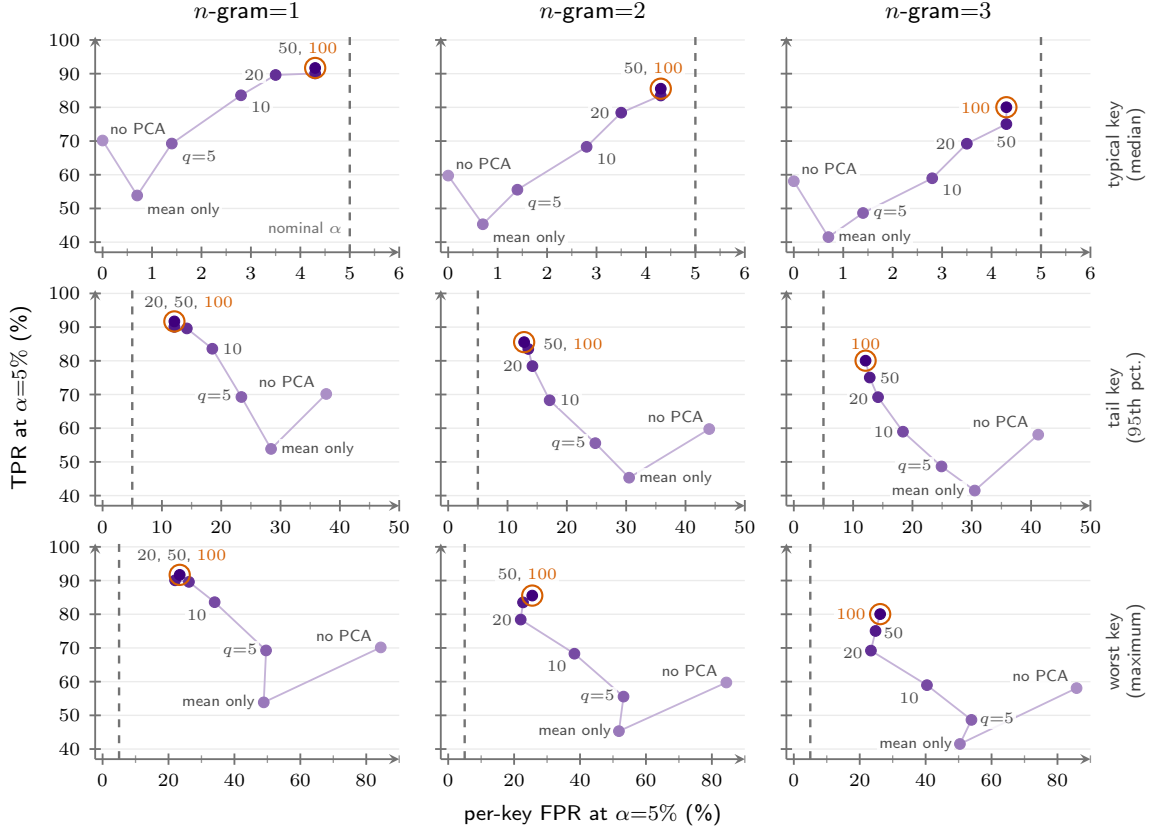}
  \caption{\textbf{Choosing $q$: detection power against three order statistics of the key bank.} The projection sweeps from \emph{no PCA} through \emph{mean only} (subtract the pool mean, remove no subspace) to $q{=}100$ removed components; the path visits the seven settings in that order and marker shade runs light to dark with $q$. \textbf{Columns} are the $n$-gram window. \textbf{Rows} are which key of the $1{,}000$-key bank the false-positive rate is read off: the typical key, the $95$th-percentile tail key and the worst key. \textbf{$y$:} TPR at $\alpha{=}5\%$ over the $1{,}200$ distilled students. \textbf{$x$:} the percentage of the $129$ clean model-temperature cells on which that key fires at the same threshold. }
  \label{fig:dehub-operating-point}
\end{figure}

\myparagraph{How many components to remove.}
We choose $q$ on clean models alone, and we read the false-positive rate of each single deployed key. Figure~\ref{fig:dehub-operating-point} sweeps $q$ from \emph{no PCA} to $q{=}100$ and reads three keys of the $1{,}000$-key bank against TPR: the typical key, the key at the $95$th percentile of the per-key rate in terms of FPR, and the worst key which has the highest FPR. Without the projection those three disagree completely, the typical key firing on no clean cell at all while the $95$th-percentile key fires on $38$ to $44\%$ of them. Removing $q{=}100$ directions closes that spread onto the nominal line and lifts TPR at $\alpha{=}5\%$ from $70.2$ to $91.7$ at $n$-gram${=}1$. A small $q$ does neither, because the leading directions carry watermark signal as well as shared task structure: the projection pays for itself only from $q{\approx}10$, after which power climbs monotonically. We deploy $q{=}100$, and App.~\ref{app:calibration-nulls} reports the per-key calibration that remains at that setting.

\FloatBarrier
\subsection{Calibration of the Rank $p$-Value}
\label{app:calibration-nulls}
\begin{figure}[h]
  \centering
  \begin{subfigure}{0.3\linewidth}
    \centering
    \includegraphics[width=\linewidth]{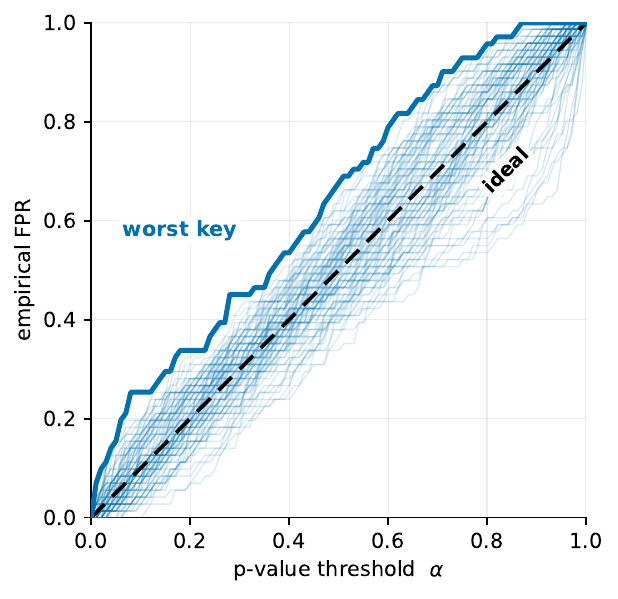}
    \caption{\method (ours)}
    \label{fig:calib-curve-realgrad}
  \end{subfigure}
  \begin{subfigure}{0.3\linewidth}
    \centering
    \includegraphics[width=\linewidth]{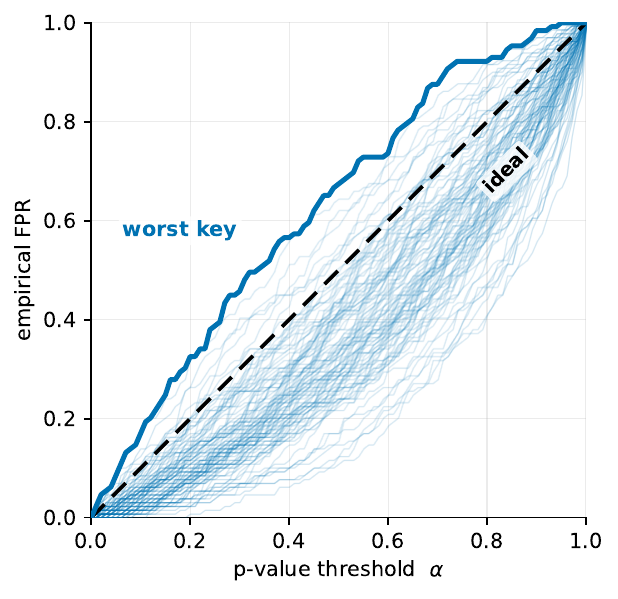}
    \caption{Black-box baseline}
    \label{fig:calib-curve-blackbox}
  \end{subfigure}
  \caption{\textbf{Nominal $p$-value thresholds match empirical false-positive rates.} Each faint curve is one of the $100$ deployed keys, reporting its empirical FPR over the $65$ held-out clean model-temperature cells at $n$-gram${=}3$ as the threshold $\alpha$ varies; the solid curve is the worst key and the dashed line is the calibrated target $\mathrm{FPR}{=}\alpha$.}
  \label{fig:calib-curve-perkey}
\end{figure}

By the exchangeability argument of App.~\ref{app:key-cal}, thresholding at $p{\le}\alpha$ should falsely accuse an $\alpha$ fraction of models that were not distilled from the protected teacher, which is what \S\ref{subsec:exp-setup} relies on when it reads operating points off the nominal $p$-value. We test it on $100$ deployed keys and $65$ clean model-temperature cells held out from the $64$ used to fit the task subspace (App.~\ref{app:calib-task-removal}), ranking each key among the full $1{,}000$-key bank. Figure~\ref{fig:calib-curve-perkey} plots the empirical rate against $\alpha$ for each deployed key, for \method (Fig.~\ref{fig:calib-curve-realgrad}) and for the black-box baseline (Fig.~\ref{fig:calib-curve-blackbox}).

\myparagraph{The per-key false-positive distribution.}
Table~\ref{tab:perkey-fpr} reports the empirical false-positive rate at $\alpha{=}5\%$ over the $1{,}000$ keys of the bank, each read across the $129$ clean model-temperature cells, after the task-subspace projection and under the same $4$-fold cross-fitting by base model as App.~\ref{app:calib-task-removal}. The typical key is calibrated and the tail is not, which is the distinction the mean over the bank hides: the median key sits at $4.3\%$ against a nominal $5\%$ in all three windows and the $90$th percentile at $9.9\%$, while the worst key reaches $23.4$ to $26.2\%$. 
That tail is bounded by the size of the pool: the \emph{floor} column is the level a perfectly exchangeable bank of this size reaches on its own (App.~\ref{app:calib-task-removal}), and against it the worst observed key is high by the \emph{worst\,/\,floor} factor, where the unprojected scores are high by ${\approx}6$.
\begin{table}[htbp]
\centering
\footnotesize
\caption{\textbf{Per-key false-positive rate at $\alpha{=}5\%$, after removing the task subspace.} Each of the $1{,}000$ bank keys is read over the $129$ clean model-temperature cells, with $4$-fold projectors cross-fitted by base model so no cell is scored by a subspace fit on its own model. All values are percentages except the last column. \emph{Floor} is the $95$th percentile of the maximum under exact uniformity at $K{=}1{,}000$ over $129$ cells: the level a perfectly exchangeable bank of this size would still reach, and the reference the worst key should be judged against.}
\label{tab:perkey-fpr}
\begin{tabular}{@{}lcccccc@{}}
\toprule
$n$-gram & median & p$90$ & p$95$ & worst & floor & worst\,/\,floor \\
\midrule
$1$ & $4.3$ & $9.9$ & $12.1$ & $23.4$ & $13.5$ & $1.74$ \\
$2$ & $4.3$ & $9.9$ & $12.8$ & $25.5$ & $13.5$ & $1.89$ \\
$3$ & $4.3$ & $9.9$ & $12.1$ & $26.2$ & $13.5$ & $1.95$ \\
\bottomrule
\end{tabular}
\end{table}

\FloatBarrier
\section{Additional Detection Results}
\label{app:additional-detection}
This section follows the sweep order of \S\ref{subsec:wild-access}, adding the operating points and factors the main tables leave out, together with two studies the main text states only in passing: the suspect-text budget behind the fragility claim of \S\ref{sec:introduction}, and the choice of probe model, together with the warm-up it is given, named in the detector setup of \S\ref{subsec:exp-setup}. The configuration of App.~\ref{app:setup} is the reference point throughout, and each row of Tab.~\ref{tab:config-experiments} names what it varies.

\subsection{Operating Points}
\label{app:roc-operating-points}
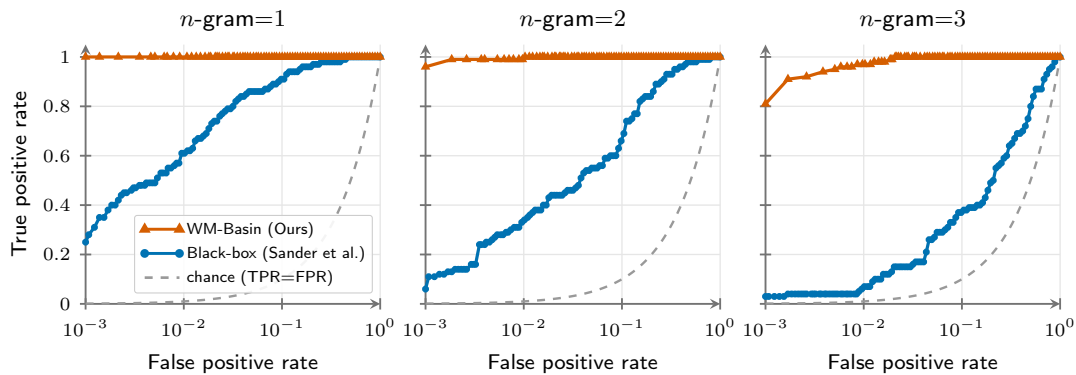
\begin{figure}[h]
  \centering
  \newcommand{\roclegend}[2]{%
  \node[anchor=south east, inner sep=2.5pt, fill=white, fill opacity=0.9,
        text opacity=1, rounded corners=1pt, draw=black!20, line width=0.4pt]
    at (rel axis cs:#1,#2) {%
    \begin{tikzpicture}[font=\figfont\fontsize{6.3}{7}\selectfont]
      \draw[detAttrcol, line width=1.1pt]
        plot[mark=triangle*, mark size=1.5pt, mark options={fill=detAttrcol, draw=detAttrcol, solid}]
        coordinates {(0cm,0cm) (0.46cm,0cm)};
      \node[anchor=base west, inner sep=0pt, yshift=-0.4ex] at (0.55cm,0cm) {\method (Ours)};
      \draw[detBBcol, line width=1.1pt]
        plot[mark=*, mark size=1.1pt, mark options={fill=detBBcol, draw=detBBcol, solid}]
        coordinates {(0cm,-0.32cm) (0.46cm,-0.32cm)};
      \node[anchor=base west, inner sep=0pt, yshift=-0.4ex] at (0.55cm,-0.32cm) {Black-box (Sander et al.)};
      \draw[detIdealcol, line width=1.0pt, dashed] (0cm,-0.64cm) -- (0.46cm,-0.64cm);
      \node[anchor=base west, inner sep=0pt, yshift=-0.4ex] at (0.55cm,-0.64cm) {chance (TPR${=}$FPR)};
    \end{tikzpicture}};%
}

\begin{tikzpicture}[/utils/exec=\figfont]
  \begin{groupplot}[
    group style={group size=3 by 1, horizontal sep=0.6cm,
      ylabels at=edge left, yticklabels at=edge left},   %
    width=0.25\textwidth, height=0.22\textwidth,
    scale only axis,
    xmode=log, log basis x=10,   %
    xmin=1e-3, xmax=1, ymin=0, ymax=1.05,
    unbounded coords=discard,   %
    xtick={1e-3,1e-2,1e-1,1}, ytick={0,0.2,0.4,0.6,0.8,1},
    xlabel={False positive rate},
    axis lines=left,
    axis line style={black!55, line width=0.7pt},
    tick style={black!55, line width=0.7pt},
    label style={font=\figfont\footnotesize},
    tick label style={font=\figfont\scriptsize},
    ylabel style={yshift=-2pt},
    xmajorgrids=true, ymajorgrids=true,
    major grid style={draw=black!10, line width=0.5pt},
    title style={font=\figfont\small, yshift=-1.0ex},
    clip=false,
  ]
  \nextgroupplot[title={$n$-gram${=}1$}, ylabel={True positive rate}]
    \addplot[detIdeal, domain=1e-3:1, samples=200, forget plot] {x};          %
    \addplot[detBB, mark size=0.7pt] table[x=fpr, y=tpr, col sep=comma]{figures_new/roc_closedmodel_by_ngram_blackbox_ng1.csv};
    \addplot[detAttr, mark size=0.9pt] table[x=fpr, y=tpr, col sep=comma]{figures_new/roc_closedmodel_by_ngram_ours_ng1.csv};%
    \roclegend{0.975}{0.04}
  \nextgroupplot[title={$n$-gram${=}2$}]
    \addplot[detIdeal, domain=1e-3:1, samples=200, forget plot] {x};          %
    \addplot[detBB, mark size=0.7pt] table[x=fpr, y=tpr, col sep=comma]{figures_new/roc_closedmodel_by_ngram_blackbox_ng2.csv};
    \addplot[detAttr, mark size=0.9pt] table[x=fpr, y=tpr, col sep=comma]{figures_new/roc_closedmodel_by_ngram_ours_ng2.csv};%
  \nextgroupplot[title={$n$-gram${=}3$}]
    \addplot[detIdeal, domain=1e-3:1, samples=200, forget plot] {x};          %
    \addplot[detBB, mark size=0.7pt] table[x=fpr, y=tpr, col sep=comma]{figures_new/roc_closedmodel_by_ngram_blackbox_ng3.csv};
    \addplot[detAttr, mark size=0.9pt] table[x=fpr, y=tpr, col sep=comma]{figures_new/roc_closedmodel_by_ngram_ours_ng3.csv};%
  \end{groupplot}
\end{tikzpicture}
  \caption{\textbf{Detection performance of the two closed-model detectors across all operating points.} ROC (log $x$, linear $y$), one panel per $n$-gram window, matched on the $100$ distillation keys $1001$ to $1100$. \textbf{TPR} is over the $100$ watermarked students (one per key, scored with its own key); \textbf{FPR} is over the same $100$ keys $\times$ the $65$ held-out innocent model-temperature cells (the evaluation half of the $129$-cell clean pool: $43$ off-the-shelf base models $\times$ $3$ sampling temperatures; $6{,}500$ null cases). Sweeping the threshold $\alpha$ traces each curve; \method (ours) is drawn in vermilion and Black-box (baseline) in blue. The grey dashed line marks chance ($\mathrm{TPR}{=}\mathrm{FPR}$).}
  \label{fig:roc-by-ngram}
\end{figure}

Figure~\ref{fig:roc-by-ngram} sweeps the detection threshold continuously over the same $T{=}1.0$ distillation arm that Tab.~\ref{tab:access-detection} reads at three fixed thresholds, on an empirical false-positive axis rather than that table's nominal $p$-values.
Both detectors are scored on one shared set of students and innocent cells here, all $100$ keys rather than the per-window cohorts the tables report, so the comparison to make is between the two curves rather than between a curve and a cell of Tab.~\ref{tab:access-detection}: \method sits at or above the black-box baseline throughout the plotted range at every window, and strictly above it everywhere below the false-positive rate at which both saturate. The baseline's curve is the read that preceded the token matching both tables now use; that matching lifts eight of the nine black-box cells of Tab.~\ref{tab:access-detection} at $5\%$ FPR, by up to $18$ points, so the separation plotted here is wider than the matched one.

\subsection{Cross-Teacher Detection}
\label{app:cross-teacher-details}
Section~\ref{subsec:wild-access} reports cross-teacher detection in Tab.~\ref{tab:cross-teacher}, whose three thresholds are empirical false-positive rates. The sampled-token baseline can only accumulate evidence at positions whose context window recurs, and student diversity orders it: gpt-oss students repeat only $65.3\%$ of their $2$-word contexts and the baseline reaches $18\%$ there, against $92.1\%$ and $93\%$ on the most repetitive Llama students. Neither tokenizer sharing, trace length nor utility separates the teachers the same way (Tab.~\ref{tab:teacher-props}), and gpt-oss gives the second-largest GSM8K gain alongside the weakest sampled-token result. \method aggregates the watermark's gradient direction over all positions instead, and stays at $96$ to $100\%$ in the $5\%$ column on every teacher.
\begin{table}[htbp]
\centering
\footnotesize
\setlength{\tabcolsep}{4pt}
\renewcommand{\arraystretch}{1.15}
\caption{\textbf{Teacher properties, student utility and detection, for the three swapped-in teachers.} \emph{DS tok} is the DeepSeek-R1-Distill tokenizer. Accuracy is mean $\pm$ standard deviation over $100$ keyed Qwen2.5-3B students, and $\Delta$ is against the un-distilled base (GSM8K $48.5\%$, Math-500 $33.8\%$). The two detection rows are the $\text{FPR}{=}5\%$ column of Tab.~\ref{tab:cross-teacher}.}
\label{tab:teacher-props}
\housetablematch{%
\begin{tabular}{l ccc}
\toprule
 & gpt-oss-120B & QwQ-32B & Llama-3.1-8B \\
\midrule
\multicolumn{4}{@{}l}{\emph{Teacher}} \\
Family / tokenizer      & MoE / $200$k native & Qwen2.5 / DS tok & Llama / native \\
Median trace chars      & $10{,}552$ & $23{,}095$ & $2{,}969$ \\
\midrule

\multicolumn{4}{@{}l}{\emph{Student utility}} \\
GSM8K (\%)              & $77.2 \pm 1.7$ & $80.8 \pm 2.7$ & $74.9 \pm 1.2$ \\
$\Delta$ GSM8K          & $+28.7$    & $+32.3$    & $+26.4$   \\
Math-500 (\%)           & $47.9 \pm 1.5$ & $52.1 \pm 1.5$ & $36.5 \pm 1.5$ \\
$\Delta$ Math-500       & $+14.1$    & $+18.3$    & $+2.7$    \\
\midrule
\multicolumn{4}{@{}l}{\emph{Detection, TPR at the $5\%$ threshold}} \\
\method                 & \hoc{96}   & \hoc{99}   & \hoc{100} \\
Black-box               & \hbc{18}   & \hbc{88}   & \hbc{93}  \\
\bottomrule
\end{tabular}%
}
\end{table}

\FloatBarrier
\subsection{Watermark Scheme}
\label{app:watermark-generality}
Table~\ref{tab:watermark-generality} replaces the teacher's TextSeal decoder with three other watermarking schemes. TextSeal, the dual-key scheme used everywhere else in the paper~\citep{sander2026textseal}, draws each token's Gumbel noise from one of two keys with probability $\alpha{=}0.1$; GumbelMax~\citep{aaronson2022gumbelmax} is the single-key Gumbel scheme; and SynthID~\citep{dathathri2024scalable} resolves each position through a tournament of depth $10$ between keyed g-values. KGW~\citep{kirchenbauer2023watermark} instead adds a fixed logit bias $\delta{=}2.0$ to a keyed half of the vocabulary ($\gamma{=}0.5$) at every position, the one scheme here that modifies the teacher's logits rather than the sampling rule. Each is generated at $n$-gram${=}1$, $2$ and $3$ and distilled into $50$ students, with all detectors reading the same ${\approx}1$M-token budget, the reference bank rebuilt per scheme and window, and everything else at the defaults of Tab.~\ref{tab:config-defaults}. 

Does \method's margin hold when the baseline is read in its strongest form? The black-box statistic can weight each scored token by the teacher's entropy at that position, concentrating the test on the high-entropy positions where the watermark is able to bias the sample at all, as in~\citep{sander2026textseal}; Tab.~\ref{tab:watermark-generality-entropy} gives the baseline that weighted read as a second row per scheme, beside the plain statistic reported in Tab.~\ref{tab:watermark-generality}. The weighting is worth $8$ to $30$ points to the baseline at $p{\le}0.05$ across all twelve scheme and window cells, yet \method still leads at every level in every cell where that level is defined. Its margin in the main body is therefore not an artefact of reading the baseline in its plain form.

\newcommand{\hoshade}[1]{\cellcolor{ourshue!#1}\na}
\newcommand{\hospan}[1]{\cellcolor{ourshue!#1}\multirow{-2}{*}{\makebox[\hcw][c]{#1}}}
\newcommand{\hnrspan}{\multirow{-2}{*}{\makebox[\hcw][c]{---}}}

\begin{table}[h]
\centering
\housesetup
\caption{\textbf{Distillation detection across watermarking schemes, with the baseline's entropy-weighted statistic.} The full form of Tab.~\ref{tab:watermark-generality}: TextSeal, the teacher's default scheme, joins the three swapped-in ones, and each watermark occupies two rows, the black-box baseline~\citep{sander2024watermarking} reading the suspect text with its plain statistic on the first and with its entropy-weighted statistic on the second. The weighting belongs to that statistic, so one \method run serves both rows and its cells span the pair. Cells are read as in Tab.~\ref{tab:watermark-generality}, and the dash marks a level the $991$-key TextSeal bank at $n$-gram${=}1$ cannot resolve.}
\label{tab:watermark-generality-entropy}
\housetablematch{%
\begin{tabular}{U}
\toprule
\househead[U]{Watermark}
TextSeal                  & \hoshade{100} & \hoshade{100} & \na           & \hb{76} & \hb{44} & \hb{28} & \hoshade{100} & \hoshade{100} & \hoshade{98}  & \hb{52} & \hb{30} & \hb{6} & \hoshade{100} & \hoshade{98}  & \hoshade{84}  & \hb{22} & \hb{8}  & \hb{2} \\
TextSeal with entropy     & \hospan{100} & \hospan{100} & \hnrspan       & \hb{96} & \hb{88} & \hb{60} & \hospan{100} & \hospan{100} & \hospan{98}  & \hb{74} & \hb{54} & \hb{40} & \hospan{100} & \hospan{98}  & \hospan{84}  & \hb{46} & \hb{24} & \hb{2} \\
\addlinespace
GumbelMax                 & \hoshade{100} & \hoshade{100} & \hoshade{100} & \hb{72} & \hb{44} & \hb{28} & \hoshade{100} & \hoshade{98}  & \hoshade{98}  & \hb{52} & \hb{32} & \hb{6} & \hoshade{100} & \hoshade{100} & \hoshade{100} & \hb{22} & \hb{10} & \hb{0} \\
GumbelMax with entropy    & \hospan{100} & \hospan{100} & \hospan{100} & \hb{96} & \hb{88} & \hb{62} & \hospan{100} & \hospan{98}  & \hospan{98}  & \hb{80} & \hb{54} & \hb{28} & \hospan{100} & \hospan{100} & \hospan{100} & \hb{46} & \hb{18} & \hb{6} \\
\addlinespace
KGW                       & \hoshade{100} & \hoshade{100} & \hoshade{100} & \hb{58} & \hb{26} & \hb{8}  & \hoshade{94}  & \hoshade{86}  & \hoshade{56}  & \hb{40} & \hb{20} & \hb{4} & \hoshade{90}  & \hoshade{68}  & \hoshade{34}  & \hb{16} & \hb{2}  & \hb{2} \\
KGW with entropy          & \hospan{100} & \hospan{100} & \hospan{100} & \hb{66} & \hb{46} & \hb{28} & \hospan{94}  & \hospan{86}  & \hospan{56}  & \hb{54} & \hb{32} & \hb{8} & \hospan{90}  & \hospan{68}  & \hospan{34}  & \hb{38} & \hb{10} & \hb{0} \\
\addlinespace
SynthID              & \hoshade{100} & \hoshade{100} & \hoshade{100} & \hb{64} & \hb{38} & \hb{8}  & \hoshade{100} & \hoshade{100} & \hoshade{98}  & \hb{44} & \hb{16} & \hb{0} & \hoshade{100} & \hoshade{96}  & \hoshade{78}  & \hb{24} & \hb{8}  & \hb{4} \\
SynthID with entropy & \hospan{100} & \hospan{100} & \hospan{100} & \hb{94} & \hb{68} & \hb{38} & \hospan{100} & \hospan{100} & \hospan{98}  & \hb{66} & \hb{38} & \hb{12} & \hospan{100} & \hospan{96}  & \hospan{78}  & \hb{42} & \hb{22} & \hb{6} \\
\bottomrule
\end{tabular}%
}
\end{table}

\FloatBarrier
\subsection{Audited Domain and System Prompt}
\label{app:ood-domain}
The audited-domain sweep of \S\ref{subsec:wild-access} reads Tab.~\ref{tab:ood-domain}; this section gives the per-domain rebuild and the exact prompts behind it. Every row only uses the prompt from the audit domain: the probe warm-up corpus, the PCA task subspace removal, the decoy bank, and the black-box baseline's per-key corpus filter are all built from that row's own questions and system prompt, never the teacher's math-teacher prompt, so neither detector inherits an advantage from the distillation domain. 

Each suspect is ranked against its own key plus the $999$ decoys of that row's bank, and $100$ students are scored in every cell. All five rows read that row's own held-out test pool, the three math-dataset rows under the shared \texttt{math\_neutral} system prompt and the other two under their own, so wording is constant inside the top block and the dataset alone separates those rows. On OpenR1-Math, the distilled dataset itself with held-out questions read under a prompt no suspect trained under, \method reaches $95\%$/$83\%$/$45\%$ when context window is 1/2/3, so a new system prompt does not by itself cost the verifier its read. 

\newtcolorbox{promptbox}[1]{%
  enhanced, breakable, sharp corners,
  colback=gray!4, colframe=black!70, coltitle=white,
  colbacktitle=black!70,
  boxrule=0.4pt, titlerule=0pt,
  left=4pt, right=4pt, top=2pt, bottom=2pt,
  before skip=3pt, after skip=3pt,
  fonttitle=\bfseries\footnotesize\sffamily,
  fontupper=\footnotesize\ttfamily,
  title={#1}
}

\tikzexternaldisable
\begin{promptbox}{Training domain (math): what the students actually learned on}
You are a math teacher. You will be given a math problem and you will solve it step by step.\\[2pt]
You will output your final solution like \textbackslash boxed\{ANSWER\}. Be sure to include relevant units within the brackets and fully evaluate arithmetic expressions.
\end{promptbox}

\begin{promptbox}{math\_neutral (OOD, closest to training)}
You are a helpful assistant. Solve the following problem step by step, then give the final answer.
\end{promptbox}

\begin{promptbox}{code (OOD)}
You are an expert programmer. Read the problem and write a complete, correct Python solution. Think step by step, then provide the final code.
\end{promptbox}

\begin{promptbox}{chat (OOD, farthest from training)}
You are a helpful, honest, and courteous assistant. Follow the user's instruction and give a clear, complete response.
\end{promptbox}
\tikzexternalenable

\FloatBarrier
\subsection{Suspect-Text Budget}
\label{app:token-budget}
\begin{figure*}[h]
  \centering
  \input{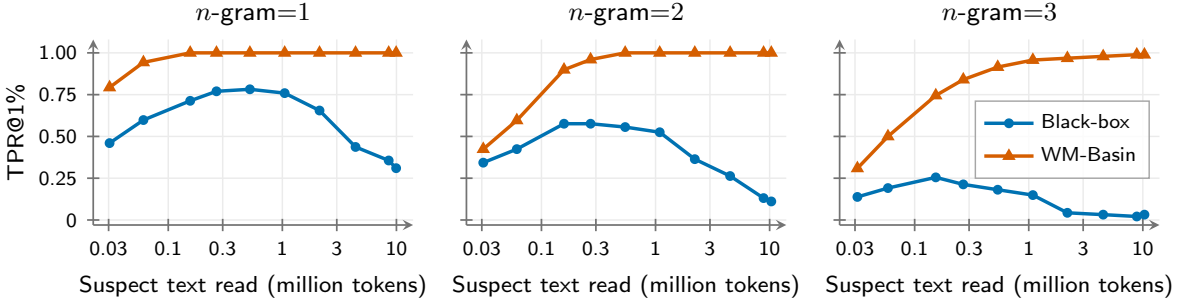}
  \caption{\textbf{Impact of suspect text budget on detection performance.} We plot True Positive Rate at $1\%$ FPR (TPR@$1\%$; share of distilled models detected at rank-$p{\le}0.01$, one model per key) against the amount of suspect text evaluated ($\log x$ scale, million tokens). 
  Both \method (ours) and \textbf{Black-box}~\citep{sander2024watermarking} are evaluated on identical generations at the same $1\%$ FPR operating point.}
  \label{fig:text-budget}
\end{figure*}

How much suspect text does a verdict need, and does reading more always help? Section~\ref{subsec:exp-setup} fixes the default budget the in-domain tables are read at. We read the same fixed pool of $100$ students, at cumulative prefixes of their generations, from $5$ to $1{,}878$ generations ($0.03$ to $10.3$M tokens supplied), holding the probe, bank, task subspace and threshold fixed, and score both detectors on every prefix at $1\%$ FPR (Fig.~\ref{fig:text-budget}). \method accumulates evidence monotonically at every window, so more text never costs the verifier anything and no stopping rule is needed: it reaches $100\%$ TPR by $0.6$M tokens at $n$-gram $1$ and $2$, and at $n$-gram${=}3$ it is within $8$ points of its ceiling by that budget and spends the twenty times more text that follows gaining the rest. The baseline is not monotone in the budget at all: it peaks within the first ${\approx}0.5$M tokens and decays afterwards, because its scored-token denominator keeps growing while deduplication caps the signal it can collect. 

\subsection{Probe Model and Warm-Up}
\label{app:probe-model}
The probe model turns text into a gradient signal. In this section, we study the impact of the probe model on the performance of \method.

\myparagraph{Choice of probe model.}
Does the verifier need a probe model from the suspect's own family? 
Table~\ref{tab:probe-model} varies only the probe $\probe$ that \S\ref{subsec:exp-setup} holds at a clean Qwen2.5-3B base, across ten checkpoints from $0.5$B to $8$B spanning four Qwen2.5 scales or post-training states, three Llama-3 checkpoints, and two families whose tokenizers sit farthest from the student's $152$k vocabulary. 
Every probe receives the same warm-up, and because each defines a different gradient coordinate space its reference bank is rebuilt with the default recipe; the suspect cohort, clean PCA pool and three $n$-gram windows stay fixed. 
Family costs the detector nothing: every probe recovers all suspects at $n$-gram${=}1$, and at $n$-gram${=}3$ and $0.1\%$ FPR the three Llama-3 probes land inside or just under the band the Qwen probes span, despite sharing no tokenizer with the student. 
The loss instead tracks how far the probe's vocabulary sits from the student's, the smallest vocabulary losing most while still staying an order of magnitude above the baseline in the same cell, and falling further behind the Qwen probes as the threshold tightens at $n$-gram${=}2$ and $3$. 
The verifier can therefore audit with whatever checkpoint it holds, preferring a nearby vocabulary.

\begin{table}[t]
\centering
\housesetup
\caption{\textbf{Effect of the probe model on detection.} The same $100$ distilled suspects are held fixed while the verifier changes the probe model and rebuilds the reference bank. Probes vary in scale, post-training, family, and tokenizer; the \textbf{Black-box} baseline~\citep{sander2024watermarking} uses no probe and is \emph{token-matched}, reading exactly the same suspect text as \method. Each cell reports True Positive Rate (TPR, \%) at FPR ($5\%, 1\%, 0.1\%$), over the suspects whose generations both detectors can score, $87/95/90$ of the $100$ keys at $n$-gram $1/2/3$. At $n$-gram${=}3$ and $0.1\%$ FPR the five Qwen2.5 probes span $69$ to $89$ and the three Llama-3 probes $67$ to $72$; vocabulary size is given beside each cross-family row.}
\label{tab:probe-model}
\housetable{%
\begin{tabular}{T}
\toprule
\househead[T]{Probe model $\probe$}
Black-box baseline (no $\probe$)        & \hb{72}  & \hb{34}  & \hb{11}  & \hb{44}  & \hb{28}  & \hb{9}  & \hb{26}  & \hb{13} & \hb{3}  \\
\midrule
Qwen2.5-3B-Base (default)               & \ho{100} & \ho{100} & \ho{100} & \ho{100} & \ho{100} & \ho{98} & \ho{100} & \ho{99} & \ho{89} \\
Qwen2.5-1.5B-Base                       & \ho{100} & \ho{100} & \ho{100} & \ho{100} & \ho{100} & \ho{99} & \ho{97}  & \ho{90} & \ho{72} \\
Qwen2.5-0.5B-Base                       & \ho{100} & \ho{100} & \ho{100} & \ho{100} & \ho{99}  & \ho{97} & \ho{99}  & \ho{96} & \ho{81} \\
Qwen2.5-3B-Instruct                     & \ho{100} & \ho{100} & \ho{100} & \ho{100} & \ho{100} & \ho{98} & \ho{100} & \ho{99} & \ho{84} \\
Qwen2.5-7B-Instruct                     & \ho{100} & \ho{100} & \ho{100} & \ho{100} & \ho{100} & \ho{96} & \ho{96}  & \ho{87} & \ho{69} \\
\hline
Llama-3.1-8B-Instruct ($128$k vocab)    & \ho{100} & \ho{100} & \ho{100} & \ho{100} & \ho{100} & \ho{92} & \ho{100} & \ho{92} & \ho{67} \\
Llama-Guard-3-8B ($128$k vocab)         & \ho{100} & \ho{100} & \ho{100} & \ho{100} & \ho{100} & \ho{94} & \ho{97}  & \ho{91} & \ho{69} \\
Llama-3.2-1B-Instruct ($128$k vocab)    & \ho{100} & \ho{100} & \ho{100} & \ho{100} & \ho{100} & \ho{93} & \ho{99}  & \ho{88} & \ho{72} \\
Gemma-2-2B-Base ($256$k vocab)          & \ho{100} & \ho{100} & \ho{100} & \ho{100} & \ho{99}  & \ho{91} & \ho{91}  & \ho{74} & \ho{57} \\
SmolLM2-1.7B-Base ($49$k vocab)         & \ho{100} & \ho{100} & \ho{100} & \ho{100} & \ho{97}  & \ho{71} & \ho{90}  & \ho{68} & \ho{34} \\
\bottomrule
\end{tabular}%
}
\end{table}

\myparagraph{Impact of warm-up.}
The probe warm-up of \S\ref{subsec:exp-setup} is one short LoRA pass on clean teacher traces, run before any key gradient is taken; 
Tab.~\ref{tab:warmup-ablation} sweeps it one knob at a time from the default, in a single scoring pass per arm. We report $n$-gram${=}3$ because it is the only window that separates the arms: at $n$-gram${=}1$ every probe in Tab.~\ref{tab:probe-model} already reads $100$ at all three levels.

Removing the warm-up costs most of the detector's power at $0.1\%$ FPR: an untrained, randomly initialised adapter keeps about a third of the default's TPR there, and the bare base model almost none. Gradients are taken with respect to the adapter parameters $\probe$ (\S\ref{subsec:closed-box}), so a random adapter still defines a usable coordinate space and retains real signal at $5\%$ FPR, whereas the bare model has no adapter to differentiate against. The warm-up therefore sharpens a signal the probe already carries rather than creating one.

Warming on prompts from different domains and datasets, i.e., chat, MMLU, GSM8K or MATH instead of the audited OpenR1-Math distribution costs $11$ to $21$ points at $0.1\%$ FPR and at most one key at $5\%$. Code is the single exception ($37\%$), and it is the warm-up corpus farthest from the reasoning traces the reference bank is built from. This is a separate question from where the verifier draws its \emph{detection} prompts, which do have to come from the audited domain or a similar one (\S\ref{subsec:verifier-prompts}).

Adapter rank and learning rate each move TPR at $0.1\%$ FPR by at most $4$ points, and epoch count by $11$; none of the three moves the $5\%$ column at all. Only divergence is fatal: at learning rate $1\mathrm{e}{-}3$ the warm-up run diverges and detection collapses at every level, a training failure the verifier can see without running the detector at all. The cheapest arms we ran are among the best at every level, so the setting to deploy is whatever converges: rank $32$ matches the default at $0.1\%$ FPR and beats it at $1\%$, and one epoch beats it at $0.1\%$. We did not run the two together, so rank $32$ at one epoch is an inference from two single-knob arms rather than a measured configuration.

\begin{table}[t]
\centering
\housesetup
\caption{\textbf{The probe warm-up is necessary, but almost none of its details are.} All rows are \method at $n$-gram${=}3$ on the same $90$ watermark keys, the same suspects, and the same suspect-text budget; only the probe warm-up changes, one knob at a time from the default. Cells are TPR (\%) at $5\%$, $1\%$, and $0.1\%$ FPR, scored on the task-neutral \emph{residual} gradients of \S\ref{subsec:task-removal}. Defaults in Tab.~\ref{tab:config-defaults}.}
\label{tab:warmup-ablation}
\housetable{%
\begin{tabular}{l @{\hgutl} *{3}{c} @{\hgutg} l @{\hgutl} *{3}{c}}
\toprule
Probe warm-up & \multicolumn{3}{c}{\method} & Probe warm-up & \multicolumn{3}{c}{\method} \\
\cmidrule(lr){2-4} \cmidrule(lr){6-8}
& \hhfpr & & \hhfpr \\
\midrule
\multicolumn{4}{@{}l}{\textit{(a) Warm-up presence}}
  & \multicolumn{4}{@{}l}{\textit{(c) Adapter rank}} \\
Trained warm-up (default)      & \ho{100} & \ho{99}  & \ho{89}
  & Rank $32$                    & \ho{100} & \ho{100} & \ho{89} \\
Untrained random adapter       & \ho{82}  & \ho{59}  & \ho{33}
  & Rank $128$ (default)         & \ho{100} & \ho{99}  & \ho{89} \\
No adapter (bare base model)   & \ho{30}  & \ho{21}  & \ho{6}
  & Rank $512$                   & \ho{100} & \ho{100} & \ho{86} \\
\cmidrule(lr){1-4} \cmidrule(lr){5-8}
\multicolumn{4}{@{}l}{\textit{(b) Warm-up corpus}}
  & \multicolumn{4}{@{}l}{\textit{(d) Learning rate}} \\
OpenR1-Math (default)          & \ho{100} & \ho{99}  & \ho{89}
  & $5\mathrm{e}{-}5$            & \ho{100} & \ho{98}  & \ho{90} \\
MATH                           & \ho{100} & \ho{97}  & \ho{76}
  & $2\mathrm{e}{-}4$ (default)  & \ho{100} & \ho{99}  & \ho{89} \\
GSM8K                          & \ho{99}  & \ho{96}  & \ho{68}
  & $5\mathrm{e}{-}4$            & \ho{100} & \ho{100} & \ho{93} \\
MMLU                           & \ho{99}  & \ho{97}  & \ho{69}
  & $1\mathrm{e}{-}3$ (diverged) & \ho{7}   & \ho{2}   & \ho{0}  \\
\cmidrule(lr){5-8}
Chat                           & \ho{99}  & \ho{94}  & \ho{78}
  & \multicolumn{4}{@{}l}{\textit{(e) Warm-up epochs}} \\
Code                           & \ho{88}  & \ho{73}  & \ho{37}
  & $1$ epoch                    & \ho{100} & \ho{99}  & \ho{92} \\
                               &          &          &
  & $4$ epochs (default)         & \ho{100} & \ho{99}  & \ho{89} \\
                               &          &          &
  & $8$ epochs                   & \ho{100} & \ho{97}  & \ho{81} \\
\bottomrule
\end{tabular}%
}
\end{table}

\section{Additional Robustness Results}
\label{app:robustness}
This section gives the protocols and weight-space analyses behind \S\ref{subsec:persistence}, followed by a robustness analysis against a paraphrasing attack that rewrites the teacher's traces before distillation.
Each modifies the student or the traces it was distilled on and re-reads it with the same probe and reference bank, at the defaults of App.~\ref{app:setup} unless this section or a caption records otherwise.

\subsection{Compression}
\label{app:persistence-compression}
This section gives the protocol behind Tab.~\ref{tab:persistence} and the weight-space displacement of every operation.
We apply per-output-channel symmetric int8, uniform group-64 int4, nf4, or $10$ to $30\%$ unstructured magnitude pruning to $100$ keyed students at each $n$-gram window, regenerating detector inputs after each operation. 
Quantized weights are dequantized to fp16 for evaluation, and pruning excludes \texttt{lm\_head} and \texttt{embed\_tokens}. A model counts as usable when its mean GSM8K accuracy is at least $0.9\times$ the distilled baseline, which int8 and prune10 clear and int4, nf4, prune20 and prune30 do not. The \emph{none} row of Tab.~\ref{tab:persistence} is the uncompressed reference rather than a compression condition, so its black-box cells were never re-read under the token matching the six operation rows use; that row therefore supports no black-box comparison, either with its own \method cells or down the column.

\begin{figure}[htbp]
\centering
\pgfplotsset{
  ppbar/.style={
    width=\linewidth, height=4.9cm,
    axis lines=left, axis line style={black!55, line width=0.6pt},
    tick style={black!55, line width=0.6pt},
    xmin=0.5, xmax=7.5, xtick={1,2,3,4,5,6,7},
    xticklabels={base,int8,prune10,prune20,int4,nf4,prune30},
    xticklabel style={rotate=45, anchor=east, font=\figfont\fontsize{6.5}{7.5}\selectfont},
    label style={font=\figfont\scriptsize},
    tick label style={font=\figfont\fontsize{7}{8}\selectfont},
    ymajorgrids=true, major grid style={draw=black!8, line width=0.5pt},
    enlarge x limits=0.08, clip=false,
    title style={font=\figfont\fontsize{7.5}{8.5}\selectfont, yshift=-3pt},
  },
}
\begin{subfigure}[b]{0.35\linewidth}
\centering
\includegraphics[width=\linewidth]{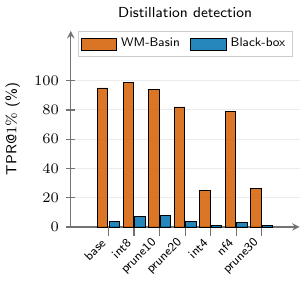}
\end{subfigure}
\hfill
\begin{subfigure}[b]{0.30\linewidth}
\centering
\includegraphics[width=\linewidth]{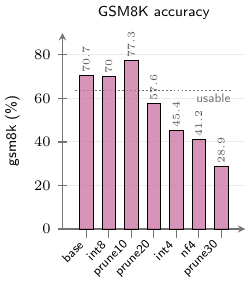}
\end{subfigure}
\hfill
\begin{subfigure}[b]{0.30\linewidth}
\centering
\includegraphics[width=\linewidth]{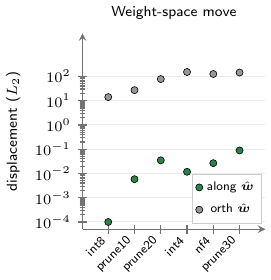}
\end{subfigure}
\caption{\textbf{Robustness to post-processing.} TextSeal, $n$-gram${=}3$, $100$ keyed students under int8, int4 and nf4 quantization and $10/20/30\%$ pruning; \emph{base} is uncompressed. \textbf{(a)} TPR at $1\%$ FPR for \textcolor{detAttrcol}{\method} and \textcolor{detBBcol}{Black-box}; \textbf{(b)} gsm8k accuracy, dotted line is the $0.9{\times}$ usable gate; \textbf{(c)} weight-space move along the watermark direction vs orthogonal residual. Details in Tab.~\ref{tab:persistence-geometry-full}.}
\label{fig:postproc-standard}
\end{figure}

Compression costs the watermark only where it costs the model. Int8 and prune10 retain both utility and strong \method detection, and the larger detection losses occur only beyond the usability gate, with nf4 and prune20 the two operations where the gradient signal outlasts utility (Fig.~\ref{fig:postproc-standard} at $n$-gram${=}3$; Tabs.~\ref{tab:persistence} and~\ref{tab:persistence-geometry-full} at all three windows). The weight-space displacement is ordered the same way: the along component of every compression operation stays below $0.11$ against orthogonal movement of $14$ to $152$, and grows with the pruning fraction. Therefore, compression does not move the model outside the watermark basin, and detection remains effective.

\begin{table}[t]
\centering
\footnotesize
\setlength{\tabcolsep}{6pt}
\renewcommand{\arraystretch}{1.15}
\caption{\textbf{Compression moves almost entirely orthogonally to the watermark direction.} Across $100$ students per window, Acc. is mean GSM8K (\%) and \emph{Usable?} denotes Acc.${\ge}0.9{\times}70.7{=}63.6$. We decompose $\Delta_{\mathrm{post}}{=}\boldsymbol{\theta}_{\mathrm{post}}{-}\boldsymbol{\theta}_{\mathrm{wm}}$ into its signed projection along the watermark key direction $\dwmhat$ (Eq.~\eqref{eq:watermark_direction}) and the orthogonal $L_2$ remainder. The remainder is a property of the compressed weights alone and is identical at the three windows; the usability verdict is read off each window's own Acc.\ and comes out the same in every row. Both are therefore printed once. The along component is at least three orders of magnitude smaller and increases in magnitude within the ordered pruning sweep.}
\label{tab:persistence-geometry-full}
\begin{tabular}{l r c rr rr rr}
\toprule
 & & & \multicolumn{2}{c}{$n$-gram${=}1$} & \multicolumn{2}{c}{$n$-gram${=}2$} & \multicolumn{2}{c}{$n$-gram${=}3$} \\
\cmidrule(lr){4-5}\cmidrule(lr){6-7}\cmidrule(lr){8-9}
Cond. & Orthogonal & Usable? & Acc. & Along $w$ & Acc. & Along $w$ & Acc. & Along $w$ \\
\midrule
none     & $0$      & \cmark & $70.7$ & $0$       & $70.7$ & $0$       & $70.7$ & $0$       \\
int8     & $14.09$  & \cmark & $72.2$ & $-0.0001$ & $71.0$ & $-0.0002$ & $70.0$ & $-0.0001$ \\
int4     & $152.07$ & \xmark & $50.4$ & $-0.0113$ & $50.8$ & $-0.0112$ & $45.4$ & $-0.0117$ \\
nf4      & $125.91$ & \xmark & $55.8$ & $-0.0291$ & $42.9$ & $-0.0281$ & $41.2$ & $-0.0266$ \\
prune10  & $27.23$  & \cmark & $77.1$ & $-0.0063$ & $76.2$ & $-0.0061$ & $77.3$ & $-0.0058$ \\
prune20  & $77.74$  & \xmark & $59.5$ & $-0.0390$ & $61.2$ & $-0.0368$ & $57.6$ & $-0.0348$ \\
prune30  & $144.85$ & \xmark & $37.8$ & $-0.1022$ & $39.6$ & $-0.0945$ & $28.9$ & $-0.0887$ \\
\bottomrule
\end{tabular}
\end{table}

\begin{table}[t]
\centering
\footnotesize
\setlength{\tabcolsep}{3.5pt}
\renewcommand{\arraystretch}{1.15}
\caption{\textbf{GRPO preserves utility with far smaller weight movement than SFT.} Every row is paired against its own cohort's pre-attack students, whose accuracy is printed as \emph{base}: the cohorts differ in size ($n$) and do not share a baseline, so $\Delta$ is the mean per-key change and not the difference of the two pooled columns beside it. GSM8K and Math-500 are percentages and $\Delta$ is in points. Move (total) is $\lVert\Delta_{\mathrm{post}}\rVert_2$, Along $w$ is its signed projection on the key direction $\dwmhat$, and Along/total is the absolute percentage share. GRPO projections are under $0.0005$ in magnitude and print as $0.000$. The SFT rows share the full $100$-key pool as their base; the GRPO rows are smaller cohorts carrying their own, which run from $0.6$ points above it to $2.8$ below, which is why Tab.~\ref{tab:continued-sft} reports accuracy as a paired change rather than an absolute.}
\label{tab:continued-sft-geometry}
\begin{tabular}{l c c rrr rrr rrr}
\toprule
& & & \multicolumn{3}{c}{GSM8K} & \multicolumn{3}{c}{Math-500} & \multicolumn{3}{c}{Move} \\
\cmidrule(lr){4-6}\cmidrule(lr){7-9}\cmidrule(lr){10-12}
Cond. & $n$-gram & $n$ & base & post & $\Delta$ & base & post & $\Delta$ & total & along $w$ & along/total \\
\midrule
reason    & $1$ & $100$ & $67.2$ & $72.0$ & $+4.77$  & $41.9$ & $40.8$ & $-1.11$ & $4.80$ & $-0.252$ & $5.3\%$ \\
reason    & $2$ & $100$ & $69.0$ & $70.0$ & $+0.96$  & $43.0$ & $39.4$ & $-3.60$ & $4.70$ & $-0.182$ & $3.9\%$ \\
reason    & $3$ & $100$ & $69.0$ & $71.3$ & $+2.37$  & $43.2$ & $40.6$ & $-2.62$ & $4.66$ & $-0.146$ & $3.1\%$ \\
\addlinespace[2pt]
multilingual & $1$ & $100$ & $67.2$ & $74.2$ & $+6.98$ & $41.9$ & $46.9$ & $+4.99$ & $2.72$ & $-0.013$ & $0.5\%$ \\
multilingual & $2$ & $100$ & $69.0$ & $73.6$ & $+4.59$ & $43.0$ & $46.7$ & $+3.68$ & $2.71$ & $-0.009$ & $0.3\%$ \\
multilingual & $3$ & $100$ & $69.0$ & $75.0$ & $+6.03$ & $43.2$ & $47.5$ & $+4.29$ & $2.70$ & $-0.007$ & $0.3\%$ \\
\midrule
rl (gsm8k, GRPO)                   & $1$ & $20$  & $67.8$ & $70.1$ & $+2.35$  & $41.3$ & $42.2$ & $+0.90$ & $0.051$ & $0.000$ & $0.0\%$ \\
rl (gsm8k, GRPO)                   & $2$ & $20$  & $69.3$ & $71.5$ & $+2.17$  & $42.5$ & $43.2$ & $+0.74$ & $0.049$ & $0.000$ & $0.0\%$ \\
rl (gsm8k, GRPO)                   & $3$ & $20$  & $68.0$ & $68.5$ & $+0.52$  & $43.4$ & $44.0$ & $+0.67$ & $0.050$ & $0.000$ & $0.0\%$ \\
\midrule
rl (K\&K, GRPO)                    & $1$ & $10$  & $66.4$ & $66.9$ & $+0.48$  & $40.9$ & $42.4$ & $+1.44$ & $0.228$ & $0.000$ & $0.0\%$ \\
rl (K\&K, GRPO)                    & $2$ & $10$  & $67.8$ & $68.7$ & $+0.93$  & $42.9$ & $43.4$ & $+0.54$ & $0.226$ & $0.000$ & $0.0\%$ \\
rl (K\&K, GRPO)                    & $3$ & $10$  & $66.2$ & $66.0$ & $-0.20$  & $43.7$ & $44.8$ & $+1.04$ & $0.241$ & $0.000$ & $0.0\%$ \\
\addlinespace[2pt]
rl (IFEval, GRPO)                  & $1$ & $10$  & $66.4$ & $65.0$ & $-1.42$  & $40.9$ & $40.3$ & $-0.58$ & $0.242$ & $0.000$ & $0.0\%$ \\
rl (IFEval, GRPO)                  & $2$ & $10$  & $67.8$ & $68.7$ & $+0.96$  & $42.9$ & $42.1$ & $-0.76$ & $0.251$ & $0.000$ & $0.0\%$ \\
rl (IFEval, GRPO)                  & $3$ & $10$  & $66.3$ & $66.1$ & $-0.26$  & $43.3$ & $41.6$ & $-1.71$ & $0.246$ & $0.000$ & $0.0\%$ \\

\bottomrule
\end{tabular}
\end{table}
\begin{table}[h]
\centering
\footnotesize
\setlength{\tabcolsep}{6pt}
\renewcommand{\arraystretch}{1.15}
\caption{\textbf{Cross-domain GRPO improves the task it optimises.} Pre, post and $\Delta$ are task accuracy in percent on greedy held-out evaluation ($400$ K\&K puzzles, $1{,}000$ IFEval prompts). Each task and window pools $10$ keys, except IFEval at $n$-gram${=}3$ ($11$). Counted over the three windows together, $10$ keys improved on K\&K and $10$ on IFEval. The same students' weight-space displacement and their off-target GSM8K and Math-500 changes are in Tab.~\ref{tab:continued-sft-geometry}.}
\label{tab:crossdomain-rl}
\begin{tabular}{l rrr rrr rrr}
\toprule
& \multicolumn{3}{c}{$n$-gram${=}1$} & \multicolumn{3}{c}{$n$-gram${=}2$} & \multicolumn{3}{c}{$n$-gram${=}3$} \\
\cmidrule(lr){2-4}\cmidrule(lr){5-7}\cmidrule(lr){8-10}
Task & Pre & Post & $\Delta$ & Pre & Post & $\Delta$ & Pre & Post & $\Delta$ \\
\midrule
K\&K   & $10.0$ & $14.4$ & $+4.4$ & $9.0$  & $13.4$ & $+4.4$ & $10.1$ & $15.2$ & $+5.1$ \\
IFEval & $18.4$ & $21.3$ & $+2.9$ & $18.3$ & $21.6$ & $+3.3$ & $18.2$ & $22.1$ & $+3.9$ \\
\bottomrule
\end{tabular}
\end{table}

\subsection{Continued Fine-Tuning and RL}
\label{app:continued-sft}
This section gives the training recipes behind Tab.~\ref{tab:continued-sft} and the weight-space displacement of every arm.
We continue training the keyed Qwen2.5-3B students of each $n$-gram window on clean data.
We then regenerate suspect text and re-measure detection, utility and geometry.
Continued SFT uses one of two clean corpora.
The \emph{reason} corpus is $1{,}900$ traces, $950$ code and $950$ science, from open-r1/Mixture-of-Thoughts~\citep{openr1}.
The \emph{multilingual} corpus is $1{,}206$ traces on math questions in eight languages~\citep{chen2024breaking}.
Both pass the think-closure and repetition filters of App.~\ref{app:distillation-details}.

SFT keeps the distillation adapter shape.
It runs $2$ epochs at global batch $16$ and maximum sequence length $8192$, on a cosine learning rate of $2{\times}10^{-5}$ with $0.05$ warmup.
Continued RL uses GRPO~\citep{shao2024deepseekmath}.
The in-domain arm optimises gsm8k with a verifiable correctness reward.
The two cross-domain arms use learning rate $5{\times}10^{-6}$ on rule-verified K\&K logic puzzles~\citep{xie2025memorization} or on RLVR-IFeval instruction following~\citep{lambert2024tulu}.
Both cross-domain rewards are all-or-nothing: K\&K requires every inhabitant to be labeled correctly in the final boxed answer, and IFEval requires the requested constraint to hold.
We measure task accuracy greedily on $400$ held-out puzzles and $1{,}000$ held-out prompts.
Detector inputs keep the standard $T{=}1.0$, top-$p{=}0.95$ sampling.

For every condition we decompose $\Delta_{\mathrm{post}}{=}\boldsymbol{\theta}_{\mathrm{post}}{-}\boldsymbol{\theta}_{\mathrm{wm}}$ against $\dwmhat$, the key-specific watermark direction of Eq.~\eqref{eq:watermark_direction}.
The Open-weight column~\citep{sander2024watermarking} is teacher-forced on the adversary's own distillation traces.
That is the ideal case, and not a read our verifier can run.
It shows that those traces stay memorised after continued training, not that the watermark survives in the weights.

Which of these updates removes the watermark?
One of the two SFT arms does, and no RL arm does (Tab.~\ref{tab:continued-sft}).
The \emph{reason} arm puts \method at its wrong-key floor at all three windows.
The \emph{multilingual} arm only degrades it, and part of the cohort is still recovered at $5\%$ FPR.
\method stays flat against each GRPO cohort's pre-attack detection at $5\%$ FPR, at every window.
The Open-weight column stays at $86$ or above in $41$ of its $45$ cells, and its four exceptions are all in the \emph{reason} arm.
The same detector scored on held-out prompts falls by $41$ to $87$ points.
Every update is nearly orthogonal to $\dwmhat$, SFT and GRPO alike.
Its projection on the watermark direction is at most $5.3\%$ of its length under SFT and $0.0\%$ under every GRPO arm (Tab.~\ref{tab:continued-sft-geometry}).
The watermark component of the weights is left in place, and under GRPO so is the text read.
Continued SFT changes what the student writes instead.
\method and the black-box baseline score that text alone, so both fall on the \emph{reason} arm at $5\%$ FPR.

\subsection{Paraphrasing}
\label{app:paraphrase}
We paraphrase only the reasoning body between \texttt{<think>} and \texttt{</think>}, preserving the tags and the final answer, then reapply the standard quality filter before distillation. A same-family (Qwen) paraphraser rewrites the traces under mild, moderate and aggressive instructions at $T{=}0.5$, $0.8$ and $1.0$ respectively, with top-$p{=}0.95$ and at most $8{,}192$ generated tokens. Each setting trains $20$ keyed students. Rewriting costs the adversary no capability: mean GSM8K accuracy exceeds $77\%$ at every strength and window against $68$ to $69\%$ for the unmodified control, and every strength stays above its matched clean student in the paired per-key comparison.

\begin{table}[t]
\centering
\housesetup
\caption{\textbf{Effect of paraphrasing teacher traces before distillation, at all three $n$-gram windows.} Every watermarked trace is rewritten at one of three strengths with a same-family (Qwen) paraphraser; \emph{none} is the unmodified control, repeated in each block because each window is a separate training set. \method (ours) and \textbf{Black-box}~\citep{sander2024watermarking} score text, while \textbf{Open-weight}~\citep{sander2024watermarking} teacher-forces the original teacher traces (\emph{orig.}) or the rewritten traces the student actually trained on (\emph{ideal}), the latter an upper bound that assumes the verifier already holds the adversary's rewritten training set. Each cell reports TPR (\%) at FPR ($5\%, 1\%, 0.1\%$) over $20$ students, except at $n$-gram${=}1$, which is scored over the $15$ of $20$ keys whose reference direction at that window is clean and whose strictest column is $0.101\%$ FPR, the finest rate that bank resolves.}
\label{tab:paraphrase}
\housetable{%
\begin{tabular}{Z}
\toprule
\househead[ZS]{Strength}
\multicolumn{14}{l}{\emph{$n$-gram${=}1$} ($15$ of $20$ keys; strictest column is $0.101\%$ FPR)} \\
none       & $69.0$ & \ho{100} & \ho{100} & \ho{100} & \hb{60} & \hb{20} & \hb{0} & \hr{100} & \hr{93}  & \hr{93} & \hr{100} & \hr{93}  & \hr{93}  \\
\addlinespace[3pt]
mild       & $77.4$ & \ho{100} & \ho{100} & \ho{100} & \hb{47} & \hb{13} & \hb{0} & \hr{93}  & \hr{87}  & \hr{60} & \hr{100} & \hr{100} & \hr{100} \\
moderate   & $80.5$ & \ho{33}  & \ho{0}   & \ho{0}   & \hb{7}  & \hb{0}  & \hb{0} & \hr{53}  & \hr{13}  & \hr{0}  & \hr{87}  & \hr{73}  & \hr{20}  \\
aggressive & $79.2$ & \ho{13}  & \ho{0}   & \ho{0}   & \hb{7}  & \hb{0}  & \hb{0} & \hr{7}   & \hr{7}   & \hr{0}  & \hr{33}  & \hr{13}  & \hr{7}   \\
\midrule
\multicolumn{14}{l}{\emph{$n$-gram${=}2$}} \\
none       & $69.2$ & \ho{100} & \ho{100} & \ho{100} & \hb{60} & \hb{25} & \hb{5} & \hr{100} & \hr{100} & \hr{100} & \hr{100} & \hr{100} & \hr{100} \\
\addlinespace[3pt]
mild       & $77.7$ & \ho{100} & \ho{85} & \ho{65} & \hb{20} & \hb{10} & \hb{5} & \hr{100} & \hr{100} & \hr{95} & \hr{100} & \hr{100} & \hr{100} \\
moderate   & $80.3$ & \ho{15}  & \ho{15} & \ho{5}  & \hb{15} & \hb{5}  & \hb{5} & \hr{55}  & \hr{20}  & \hr{0}  & \hr{100} & \hr{95}  & \hr{85}  \\
aggressive & $80.2$ & \ho{15}  & \ho{10} & \ho{5}  & \hb{5}  & \hb{5}  & \hb{0} & \hr{5}   & \hr{5}   & \hr{0}  & \hr{55}  & \hr{40}  & \hr{10}  \\
\midrule
\multicolumn{14}{l}{\emph{$n$-gram${=}3$}} \\
none       & $68.4$ & \ho{100} & \ho{100} & \ho{85} & \hb{35} & \hb{10} & \hb{0} & \hr{100} & \hr{100} & \hr{100} & \hr{100} & \hr{100} & \hr{100} \\
\addlinespace[3pt]
mild       & $77.8$ & \ho{90} & \ho{70} & \ho{25} & \hb{10} & \hb{5} & \hb{0} & \hr{100} & \hr{100} & \hr{95} & \hr{100} & \hr{100} & \hr{100} \\
moderate   & $81.2$ & \ho{5}  & \ho{0}  & \ho{0}  & \hb{5}  & \hb{0} & \hb{0} & \hr{55}  & \hr{25}  & \hr{15} & \hr{100} & \hr{95}  & \hr{80}  \\
aggressive & $79.5$ & \ho{0}  & \ho{0}  & \ho{0}  & \hb{0}  & \hb{0} & \hb{0} & \hr{10}  & \hr{0}  & \hr{0}  & \hr{55}  & \hr{25}  & \hr{10}  \\
\bottomrule
\end{tabular}%
}
\end{table}

Each $n$-gram window is a block of Tab.~\ref{tab:paraphrase}. Mild rewriting leaves the watermark readable from text: \method holds $100$, $100$ and $90\%$ TPR at $5\%$ FPR at $n$-gram${=}1$, $2$ and $3$, against $47$, $20$ and $10\%$ for the black-box baseline. Moderate and aggressive rewriting bring both text detectors down together, to at most $33\%$ TPR at $5\%$ FPR for \method and $15\%$ for the baseline. What is left then depends on which traces the open-weight read~\citep{sander2024watermarking} is teacher-forced on. On the original teacher traces, the set a verifier actually holds, it falls with the text detectors ($53$, $55$ and $55\%$ at $5\%$ FPR under moderate rewriting). On the rewritten traces the student trained on, it recovers the moderate arm almost in full ($87$, $100$ and $100\%$ at $5\%$ FPR, and $95\%$ at $1\%$ FPR at the two longer windows). The aggressive arm is gone under either read, at best $33$ to $55\%$ at $5\%$ FPR. Moderate rewriting therefore leaves the watermark in the weights and only hides it from the traces the verifier can obtain.

\FloatBarrier

\end{document}